\pdfoutput=1

\documentclass[11pt]{article}

\usepackage[]{acl}

\usepackage[algo2e]{algorithm2e} 
\usepackage{algorithm, algorithmic}

\usepackage{times}
\usepackage{latexsym}
\usepackage{booktabs}
\usepackage{amsmath,amsfonts}
\usepackage{multirow}
\usepackage{arydshln}
\usepackage{colortbl}
\usepackage{booktabs}
\definecolor{darkergreen}{rgb}{0.0, 0.5, 0.0}
\definecolor{darkerred}{rgb}{0.5, 0.0, 0.0}
\usepackage[framemethod=TikZ]{mdframed}
\usepackage[T1]{fontenc}
\definecolor{revised}{RGB}{0, 0, 255}

\usepackage[utf8]{inputenc}

\usepackage{microtype}

\newcommand\blfootnote[1]{%
  \begingroup
  \renewcommand\thefootnote{}\footnote{#1}%
  \addtocounter{footnote}{-1}%
  \endgroup
}

\title{~\includegraphics[height=25pt]{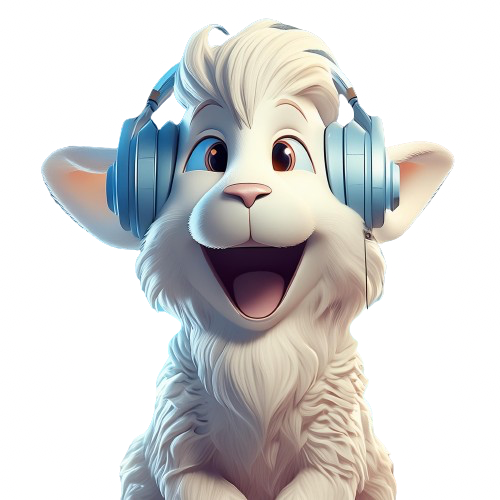} GAMA: A Large Audio-Language Model with Advanced Audio Understanding and Complex Reasoning Abilities}

\author{
    Sreyan Ghosh$^{1*}$,
    Sonal Kumar$^{1*}$,
    Ashish Seth$^{1}$,
    Chandra Kiran Reddy Evuru$^{1}$,\\
    \bf Utkarsh Tyagi$^{1}$,
    \bf S Sakshi$^{1}$,
    \bf  Oriol Nieto$^{2}$,
    \bf  Ramani Duraiswami$^{1}$,
    \bf Dinesh Manocha$^{1}$ \\
    $^{1}$University of Maryland, College Park, USA $^{2}$Adobe, USA \\
    \texttt{\{sreyang,sonalkum,dmanocha\}@umd.edu} \\ 
    Project: \url{https://sreyan88.github.io/gamaaudio/}
}

\begin{document}
\maketitle
\begin{abstract}

Perceiving and understanding non-speech sounds and non-verbal speech is essential to making decisions that help us interact with our surroundings. In this paper, we propose \texttt{GAMA}, a novel \underline{\textbf{G}}eneral-purpose Large \underline{\textbf{A}}udio-Language \underline{\textbf{M}}odel (LALM) with \underline{\textbf{A}}dvanced Audio Understanding 
and Complex Reasoning Abilities. We build \texttt{GAMA} by integrating an LLM with multiple types of audio representations, including features from a custom Audio Q-Former, a multi-layer aggregator that aggregates features from multiple layers of an audio encoder. We fine-tune \texttt{GAMA} on a large-scale audio-language dataset, which augments it with audio understanding capabilities. Next, we propose \texttt{CompA-R} (Instruction-Tuning for \underline{\textbf{Comp}}lex \underline{\textbf{A}}udio \underline{\textbf{R}}easoning), a synthetically generated instruction-tuning (IT) dataset with instructions that require the model to perform complex reasoning on the input audio. We instruction-tune \texttt{GAMA} with \texttt{CompA-R} to endow it with complex reasoning abilities, where we further add a soft prompt as input with high-level semantic evidence by leveraging event tags of the input audio. Finally, we also propose \texttt{CompA-R}-\textit{test}, a human-labeled evaluation dataset for evaluating the capabilities of LALMs on open-ended audio question-answering that requires complex reasoning. Through automated and expert human evaluations, we show that \texttt{GAMA} outperforms all other LALMs in literature on diverse audio understanding tasks by margins of 1\%-84\%. Further, GAMA IT-ed on \texttt{CompA-R} proves to be superior in its complex reasoning and instruction following capabilities.

\blfootnote{${^*}$ Co-leads with equal contribution.}

\end{abstract}

\section{Introduction}

Large Language Models (LLMs) possess impressive abilities to understand and reason about the world through language~\cite{zhao2023survey}. While spoken language understanding tasks, like automatic speech recognition, have had a long history of benefiting from language comprehension with (L)LMs~\cite{watanabe2018espnet,hu2024large}, the ability to improve the perception and understanding of non-speech sounds and non-verbal speech through language has been less explored (from hereon we refer to these kinds of audios or sound as ``audio'' in the paper). Beyond visual and language perception, the ability to understand audio is unarguably important and necessary for autonomous agents to interact with the world.

\begin{figure}[t]
    \centering
    \includegraphics[width=\columnwidth]{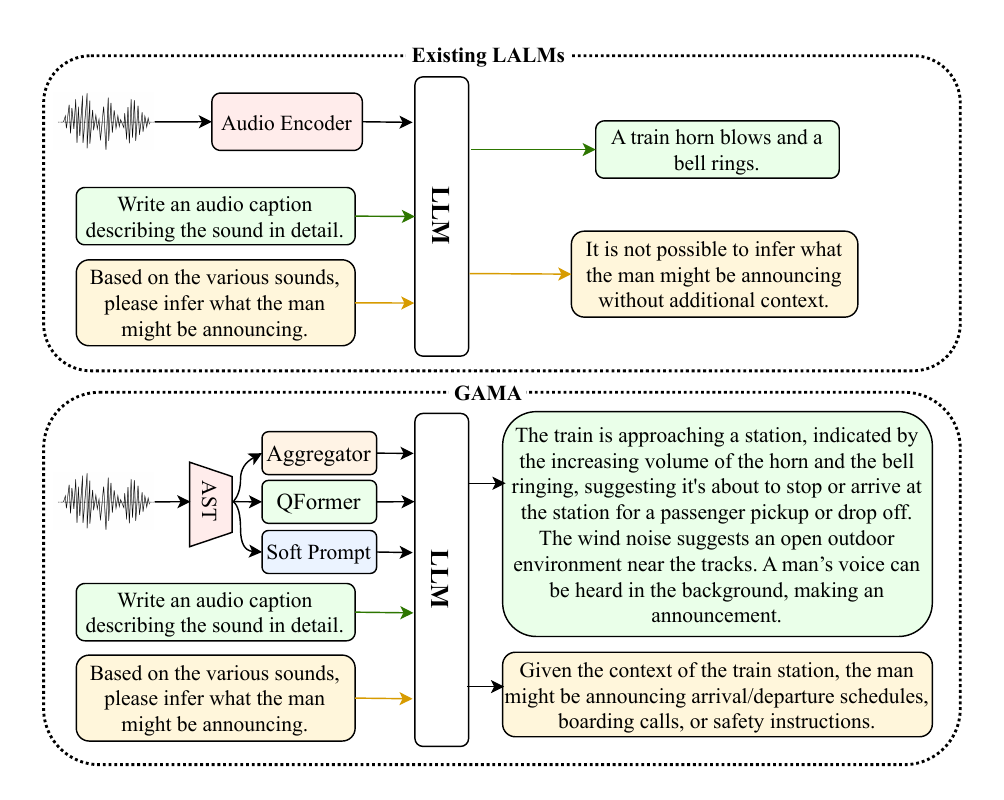}
    \caption{\small Comparison of existing LALMs (LTU by~\citet{gong2024listen} here) and \texttt{GAMA}. With improved audio understanding abilities (via diverse audio feature integration) and training on our proposed \texttt{CompA-R}, \texttt{GAMA} can provide more detailed captions of input audio and is also able to answer questions regarding it that demand complex reasoning.}
    \label{fig:first_page}
    \vspace{-1mm}
\end{figure}

Contrastive Language Audio Pre-training (CLAP)~\cite{elizalde2023clap} was one of the first Audio-Language Models (ALM) to improve audio understanding through a language interface. Following this, several attempts have been made to improve CLAP and its reasoning abilities~\cite{ghosh2024compa}. On the other hand, \citeauthor{deshmukh2023pengi} propose Pengi, a pre-trained decoder-only LLM coupled with an audio-encoder, that can solve all kinds of audio tasks by framing them as open-ended text-generation tasks. Similarly, Large Audio Language Models (LALMs) like LTU~\cite{gong2024listen} and SALMONN~\cite{tang2024salmonn} follow a similar architecture and attempt to solve audio tasks by empowering the model with instruction following capabilities~\cite{wei2022finetuned}. Specifically, all audio tasks are first framed in instruction-response pairs. The model is then fine-tuned on these pairs to learn audio reasoning and, thereby, instruction following. As an emergent ability, these models also show remarkable capabilities in open-ended question answering by reasoning over the input audio. However, two significant problems still persist: (1) All these models employ simple connection modules between the audio encoder and the language decoder to enable the latter with audio understanding capabilities. This hinders comprehensive multimodal connection and alignment, thereby increasing the risk of hallucinations and leading to suboptimal performance~\cite{liu2023improved}. (2) Complex reasoning with LALMs is still under-explored. While these models excel at audio event detection (in various forms like captioning, event classification, etc.) and information-seeking questions (e.g., close-ended audio questions like \textit{``How many birds are squawking?''}), they fail to provide a faithful response for questions involving complex reasoning like \textit{``Identifying the context of laughter and its relationship with the automotive sounds in the recording. Draw a conclusion on the possible scenario occurring.''}. We define complex reasoning for LALMs in Section~\ref{subsubsec:compar} and show examples in Fig.~\ref{fig:first_page} and Fig.~\ref{fig:qa_example1}.
\vspace{1mm}

{\noindent \textbf{Main Contributions.}} Our primary contributions are as follows:

\begin{itemize}
    \item \textbf{A Novel LALM.} We introduce \texttt{GAMA}, an LALM with advanced audio understanding and complex reasoning abilities. To improve audio perception and understanding abilities, we propose integrating an LLM with multiple types of audio features that encode diverse aspects of information about the input audio. Specifically, we couple the output features from an Audio Q-Former and an Audio Spectrogram Transformer (AST)~\cite{gong2021ast}, where the AST is further equipped with an \textit{aggregation module}. While the Audio Q-Former possesses impressive semantic generalization capabilities~\cite{li2023blip}, the AST possesses strong knowledge of surface-level audio properties. Additionally, inspired by the fact that different layers in audio models learn audio information at different scales~\cite{singla2022audio}, the aggregation module aggregates the features from multiple layers of AST, which helps encode diverse knowledge. Both representations are passed through MLP layers that connect these features into the word embedding space before adding them as the prefix. As a result, \texttt{GAMA} possesses improved audio understanding capabilities by moving away from the simple coupling of audio encoders and linear layers commonly employed as connection modules to align the audio and textual modalities, which generally suffer from comprehensive multimodal alignment~\cite{liu2023improved}. \texttt{GAMA} is first fine-tuned on a large-scale audio-language corpus, and the resulting model outperforms all other models on standard audio and music understanding benchmarks.
    \item \textbf{A Novel Instruction Tuning Dataset.} To endow an LALM with complex reasoning abilities, we propose \texttt{CompA-R}, a dataset synthetically generated with multi-aspect information and human-written in-context examples. Specifically, we prompt GPT to synthesize an instruction-response pair by guiding it with various metadata related to the audio.

    \item \textbf{A Novel Evaluation Dataset.} To evaluate an LALM's complex reasoning abilities, we develop \texttt{CompA-R}-\textit{test}, a human-labeled benchmark. Specifically, \texttt{CompA-R}-\textit{test} evaluates an LALM on open-ended AQA that demands complex reasoning over the audio. \texttt{GAMA-IT} (\texttt{GAMA} fine-tuned on \texttt{CompA-R}) shows significant improvements on \texttt{CompA-R}-\textit{test} over all other baselines from literature.
    
\end{itemize}

\begin{figure*}[t]
    \centering
    \includegraphics[width=\textwidth]{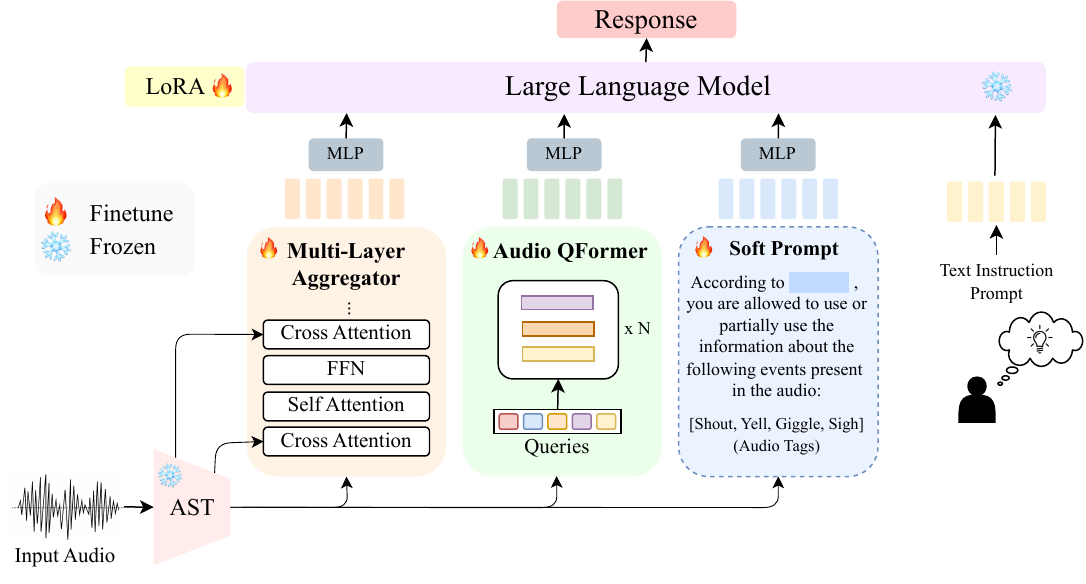}
    \vspace{-1.75em}
    \caption{\small Illustration of \texttt{GAMA}. The \texttt{GAMA} architecture involves integrating various audio features with a text-only (pre-trained) LLM. The audio features are added as a prefix to the text instruction (by the user), and the LLM responds using a text output. We feed \texttt{GAMA} with 3 different types of audio features: \textcircled{\raisebox{-0.9pt}{1}} The Multi-Layer Aggregator that takes as input features from the last and multiple other layers of the AST and calculates cross-attention to output a feature that encodes diverse surface features from audio. \textcircled{\raisebox{-0.9pt}{2}} The pre-trained Audio Q-Former takes as input the last layer features of AST and outputs a feature that encodes the audio into a semantically rich and generalized space. \textcircled{\raisebox{-0.9pt}{3}} After fine-tuning, we instruction-tune this model on our proposed \texttt{CompA-R} and further feed the model with a soft prompt with audio tags from AST (with an added classification head). This additional information serves as high-level semantic knowledge to improve complex reasoning abilities.}
    \label{fig:main_diag}
\end{figure*}

\section{Related Work}
\label{sec:related}

{\noindent \textbf{Large Multi-Modal and Audio-Language Models.}} Prior to the exploration of LLMs as efficient reasoners, encoder-based multi-modal language models, trained to learn a shared space between language and other modalities, have shown great promise. For example, CLAP, inspired by CLIP~\cite{radford2021learning} in vision, showed state-of-the-art performance on audio-language tasks like retrieval, zero-shot classification, etc.

LLMs pre-trained at an incredible scale with the next token prediction objective implicitly compress world knowledge in their parameters~\cite{zhao2023survey}. These models learn general-purpose representations, which can then be \textit{aligned} with the desired response characteristics~\cite{zhang2023instruction}. Instruction Tuning (IT), the process of fine-tuning an LLM with instruction-response pairs, has proved to be one of the most popular forms of alignment. Recent work shows that LLMs can also be instruction-tuned for multi-modal alignment. LLaVa~\cite{liu2024visual}, a pioneering work on multi-modal vision-language alignment, showed that fine-tuning an LLM on visual instruction-response pairs with additional vision features as prefix can endow the model with visual reasoning and understanding abilities. Several works following LLaVa improve aspects of LVLMs and have achieved impressive performance on several vision-language tasks~\cite{zhang2024mm}. On the other hand, LALMs like LTU and SALMONN showed impressive performance on several audio-language tasks by reasoning over the audio. Though these models extensively evaluate several closed- and open-ended tasks, their ability to perform complex reasoning is largely under-explored.

{\noindent \textbf{Instruction Tuning and Complex Reasoning.}}  IT-based alignment has also shown significant improvements for LLMs on Natural Language Understanding tasks, unlocking impressive capabilities~\cite{bubeck2023sparks}, suggesting that fine-tuning is key to building and improving LLM-based agents. Very recently, ~\citep{xu2024wizardlm} and ~\citep{cui2024adainstruct} show that well-curated IT data can improve various reasoning capabilities in LLMs, like logical, mathematical, complex reasoning, etc. More specifically, IT teaches LLMs better and more effective methods to reason about a problem, presented in the input instruction (like step-by-step reasoning~\cite{kojima2022large}).
\section{Methodology}
\label{sec:method}
In the next sub-sections, we first describe the \texttt{GAMA} architecture and its components in detail, followed by fine-tuning \texttt{GAMA} on audio-language pairs, \texttt{CompA-R} creation, and instruction-tuning \texttt{GAMA} on \texttt{CompA-R}.

\subsection{\texttt{GAMA} Architecture}

Fig.~\ref{fig:main_diag} illustrates the architecture of \texttt{GAMA}. \texttt{GAMA} builds on the same base architecture proposed in prior works~\cite{gong2024listen} but introduces several novel components for improving audio perception. More specifically, we feed the pre-trained LLM with features from multiple audio encoders, including a pre-trained Audio-Q-Former and a pre-trained AST that encode diverse audio knowledge. Additionally, unlike prior work, we do not just use the last layers of the AST but couple it with a multi-layer aggregator that takes features from multiple layers as input and outputs a feature that is aware of various low-level and high-level properties of the input audio. Finally, to endow the model with effective complex reasoning abilities, we employ AST again to extract high-level semantic knowledge, i.e., audio event tags, as supplementary information.

\subsubsection{Audio Spectrogram Transformer (AST)}
Audio Spectrogram Transformer (AST), was one of the first attempts to model audio signals with a pure Transformer network. We employ an AST model fine-tuned on the AudioSet dataset. AST has been employed as an audio encoder and a feature extractor in a wealth of prior works due to its high informativeness~\cite{gong2023contrastive, gong2024listen}. To extract the last-layer features, we drop the audio classification head and employ it only for event classification for soft prompts.

\subsubsection{Audio Q-Former}
{\noindent \textbf{Motivation.}} Our primary goal is to integrate \texttt{GAMA} with an audio encoder that possesses strong semantic generalization capabilities for any input audio. Prior work has extensively explored CLAP-style training for learning audio-language encoders. However, other methods and architectures have rarely been explored. As a more powerful alternative, we explore the Q-Former architecture proposed by~\citep{li2023blip}.
\vspace{0.5mm}

{\noindent \textbf{Architecture.}} The architecture of our Audio Q-Former is based on the Querying Transformer proposed in ~\citet{li2023blip}, which is initialized from BERT~\cite{devlin2018bert} and has $Q$ querying tokens. We employ AST as the audio encoder (in place of the ViT-based vision encoder) and keep the rest of the architecture the same. Similar to the original implementation, we train the model in two stages. For the first stage, we solve three tasks, namely the Audio-Text Matching loss, the Audio-Grounded Text Generation loss, and the Audio-Text Contrastive Learning loss. For the second stage, we employ LLaMa-2~\textsubscript{7B} as the language decoder and solve the language-modeling loss. For training, we use 2.5M+ audio-caption pairs (detailed in Section~\ref{sec:dataset}). For architectural details, we refer our readers to~\citet{li2023blip}. 

{\noindent \textbf{Training with Caption Augmentation.}} Additionally, due to the lack of large-scale audio caption pairs, we adopt a \textit{caption-augmentation} methodology to augment the existing audios with diverse additional captions. More specifically, we instruct an LLM to generate  $k$ rewrites of the original caption. We employ two different prompts that rewrite the input caption with two different objectives:

{\noindent \textbf{Prompts.}} For Prompt 1, our primary aim is that the resultant rewrite should describe each acoustic event in the caption similarly but more vividly. These augmentations help the model learn various distinctive characteristics of the audio concepts corresponding to the acoustic events. For Prompt 1, our primary aim is such that the resultant rewrite should describe each acoustic event in the caption differently from the original caption. These augmentations aid the model in understanding the diverse linguistic expressions that can describe a single audio concept. We show examples below: (more examples in Table~\ref{tab:aug_examples_prompt_train_1}):

\begin{mdframed}[linewidth=1pt, linecolor=black, leftmargin=1pt, rightmargin=1pt, innerleftmargin=10pt, innerrightmargin=10pt, innertopmargin=4pt, innerbottommargin=2pt, backgroundcolor=gray!20, roundcorner=5pt]
\textit{\textbf{(1) Original Caption:}} Someone made a cool vocal for a dubstep track.

\noindent\textit{\textbf{(1) Rewritten Caption by Prompt 1:}} A captivating vocal performance ignites the dubstep track, delivering a hypnotic and enthralling sound that reverberates through the air.

\noindent\textit{\textbf{(1) Rewritten Caption by Prompt 2:}} The dubstep track features a slick, stylish vocal performance that adds a layer of sophistication to its heavy beats and basslines.

\noindent\textit{\textbf{(2) Original Caption:}} Someone eating crisps and talking.

\noindent\textit{\textbf{(2) Rewritten Caption by Prompt 1:}} Crunchy crisps mingle with the sound of a lively conversation, creating a cozy and intimate atmosphere.

\noindent\textit{\textbf{(2) Rewritten Caption by Prompt 2:}} The crunch of crisps and the rustle of papers create a cozy, intimate atmosphere, accompanied by the gentle hum of a conversation.

\end{mdframed}

During training, for each audio sample, we choose the original caption with a probability $p = 0.4$ or one of the rewritten versions (with a probability $1-p$), where each rewritten caption has an equal probability of selection. Both instrcutions are provided in Appendix~\ref{sec:prompts}. We employ LLaMa-2-13B~\cite{touvron2023llama} with human-written in-context examples. We randomly sample 5 in-context examples from a collection of 50.

\subsubsection{Multi-Layer Aggregator}

{\noindent \textbf{Motivation.}} To extract additional details about the input audio, we devise a multi-layer aggregator that integrates multi-level hidden features of the pre-trained AST. Although AST has a global reception field in all layers, different layers learn auditory information at different scales~\cite{singla2022audio}, i.e., the middle layers encode more generic features (e.g., basic sounds, textures), while deeper layers capture high-level concepts (e.g., speech intonations, complex sound patterns). By aggregating these features, the multi-layer aggregator outputs features that encode a more holistic and fine-grained understanding of the audio. Thus, our multi-layer aggregator makes fine-grained auditory knowledge more likely to be learned while training.

{\noindent \textbf{Architecture.}} Our multi-layer aggregator is a transformer-style network consisting of two transformer layers for aggregating the hidden features of the audio encoder. Given the hidden features $A_j$ and $A_k$ from the middle layers in the audio encoder, the aggregation module uses two blocks to sequentially integrate the former two features with the last layer feature $A_i$. Each block $\mathcal{B}$ is composed of self-attention, cross-attention, and Feed-forward network (FFN) arranged in a sequential manner. Finally, the output features $\bar{A}$ is generated as follows,

\begin{equation}
\small
\bar{A}=\mathcal{B}_2\left(\mathcal{B}_1\left(A_i ; A_j\right) ; A_k\right)
\end{equation}
\vspace{-0.75em}
\begin{equation}
\small
\mathcal{B}(X ; Y)=\operatorname{FFN}(\operatorname{Cross-Attn}(\operatorname{Attn}(X), Y)) .
\end{equation}

In practice, we employ j = 4 and k = 8 from AST as our input to the multi-layer aggregator.

\subsubsection{Soft Prompt}
\label{subsubsec:soft_prompt}

{\noindent \textbf{Motivation.}} Though models like AST and Audio Q-Former have shown much promise in audio tasks, a major problem still exists: real-world audio generally has multiple and overlapping acoustic events, and understanding all such events from model features proves to be inherently complex~\cite{ghosh2024compa}. This eventually leads to sub-optimal performance for complex reasoning, where the explicit knowledge of \textit{plausible} acoustic events in the audio can improve model responses. Thus, to improve fine-grained audio perception capabilities, we augment \texttt{GAMA} with high-level semantic understanding of the input audio. To do this, we employ an off-the-shelf audio model to extract high-level semantic knowledge, i.e., audio event tags, as supplementary information. However, as audio event classification is not a solved problem, errors in tag predictions are inevitable. Thus, to mitigate the potential adverse effects of inaccurate predictions, we are inspired by prompt tuning to introduce a soft prompting technique that enables the model to utilize the embedded tags within the instructions adaptively.

{\noindent \textbf{Architecture.}}  Fig.~\ref{fig:main_diag} shows an example of how we design our soft prompt together with an instruction. Specifically, we construct a fixed instruction template where we add the audio event tags along with the soft prompt, where the soft prompt is a trainable vector. In contrast to standard prompt tuning, where the model activations are generally steered towards completing the task for which the prompt is optimized, in our version the direction is specified by a tailored input sentence, “According to <hint>, you are allowed to use or partially use the following tags:”, and “<hint>” will be replaced by the soft prompt. This design allows us to select valuable information from tags adaptively rather than serving a specific task, as seen in standard prompt tuning methods. We only employ the soft prompt in the instruction tuning for complex reasoning step and not in the fine-tuning step. We provide a rationale in Appendix~\ref{subsec:soft_appendix}.
 
\subsubsection{Connection Module}
\label{subsubsec:connect_module}

We employ a multi-layer perceptron (MLP) to connect audio features into the word embedding space.  All features are passed through separate MLP layers before being added as prefixes to word embeddings of the text instruction prompt.

\subsection{\texttt{CompA-R}}
\label{subsubsec:compar}

\begin{figure*}[t]
    \centering
    \includegraphics[width=\textwidth]{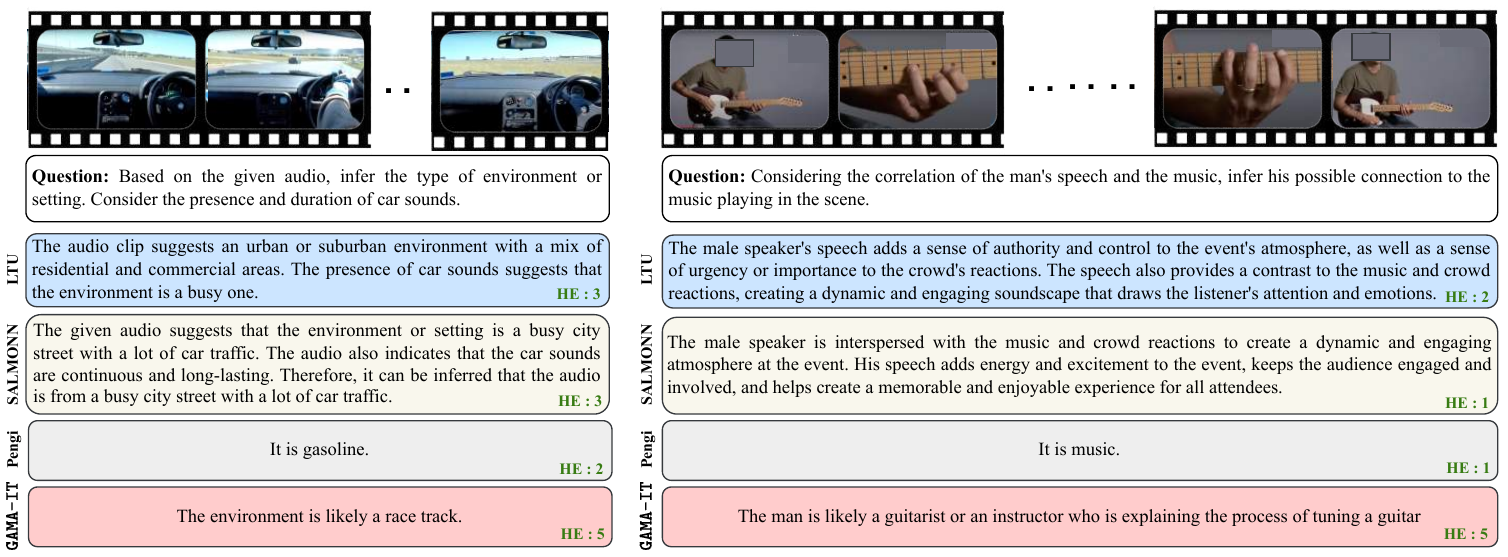}
    \caption{\small Qualitative comparison of \texttt{GAMA} with other baselines on instances from \texttt{CompA-R}-\textit{test}. Both instances challenge an LALM with a question about the input audio that requires advanced understanding and complex reasoning regarding the audio and its individual events. \texttt{GAMA} is able to provide faithful and accurate responses through improved audio perception and reasoning capabilities. \textcolor[HTML]{006633}{HE} refers to scores assigned by human evaluators. Note that the video is only provided for illustration purposes and not provided as input to the LALM. More examples here: https://gamaaudio.github.io/gamaaudio/}
    \label{fig:qa_example1}
\end{figure*}

{\noindent \textbf{Motivation.}} We define complex reasoning as the capability of an LALM to understand the input audio, every individual acoustic event in the audio, and reason the corresponding scene in which the audio might have occurred, such that it can infer nuanced relationships between them and its underlying contexts, thereby enabling it to draw sophisticated conclusions. We design \texttt{CompA-R} with the primary goal of endowing LALMs with complex reasoning abilities. We are motivated by the primary finding that current SOTA LALMs can only perform well in prompts that require describing the audio (e.g., \textit{Describe the audio}) or reasoning-based prompts where identifying the acoustic events present in the audio would suffice for a faithful response (e.g., \textit{What type of video can this audio be used for dubbing?}). However, when posed with complex reasoning questions, these models often hallucinate or fail to provide a faithful response (see Fig.~\ref{fig:qa_example1}). Inspired by a wealth of prior work that shows how IT on well-curated datasets can align model behaviors for the execution of novel skills like reasoning and complex problem solving~\cite{xu2024wizardlm}, we propose a systematic multi-stage pipeline to synthesize instruction-response pairs for \texttt{CompA-R}. \texttt{CompA-R} trains a model to engage in complex reasoning by querying it with instructions that cannot be directly inferred by identifying individual audio events and would require analyzing each event and its context in relation to other scene elements and world knowledge.

{\noindent \textbf{Synthesis Pipeline.}} We employ the AudioSet-strong subset to synthesize \texttt{CompA-R}. Our data synthesis pipeline consists of 3 stages: \textit{\textbf{i) Caption Generation.}} To generate a caption that is aware of both the audio and the visual scene, we feed GPT-4 with multiple types of information about the audio and its corresponding video. These include a caption of the middle frame of the video generated using BLIP-2~\cite{li2023blip}, objects in the frame identified using Grounding DINO~\cite{liu2023grounding}, image labels for the frame using the ImageNet~\cite{deng2009imagenet} ontology obtained from CLIP, environment context using PlaceCNN~\cite{zhou2017places}, caption of the audio obtained using RECAP~\cite{10448030} and audio event tags using the AudioSet ontology obtained from AST. Finally, we prompt GPT-4 to aggregate these descriptions into a comprehensive caption. \textit{\textbf{ii) Dataset Synthesis.}} We pass the generated caption together with the ground-truth acoustic event information and their corresponding time slices to GPT-4. We prompt GPT-4 with 3 human-written exemplars (which are randomly sampled from a pool of 50 exemplars) to synthesize an instruction-response pair. The exemplars and prompt are designed such that the synthesized instructions demand complex reasoning. We synthesize a total of 25000 instruction-response pairs. \textit{\textbf{iii) Human Verification.}} We discard instructions due to untended noise and hallucinations. We, the authors of this paper, manually verify a subset of \texttt{CompA-R} corresponding to 500 unique audios for creating the test set, i.e., \texttt{CompA-R}-\textit{test}. The remainder of the synthesized dataset is used as the training set. We describe the process and annotation details further in Appendix~\ref{sec:annotation_compar}. This finally led to 200,234 unique pairs in training and 1,561 in testing.

\begin{figure}[t]
    \centering
    \includegraphics[width=\columnwidth]{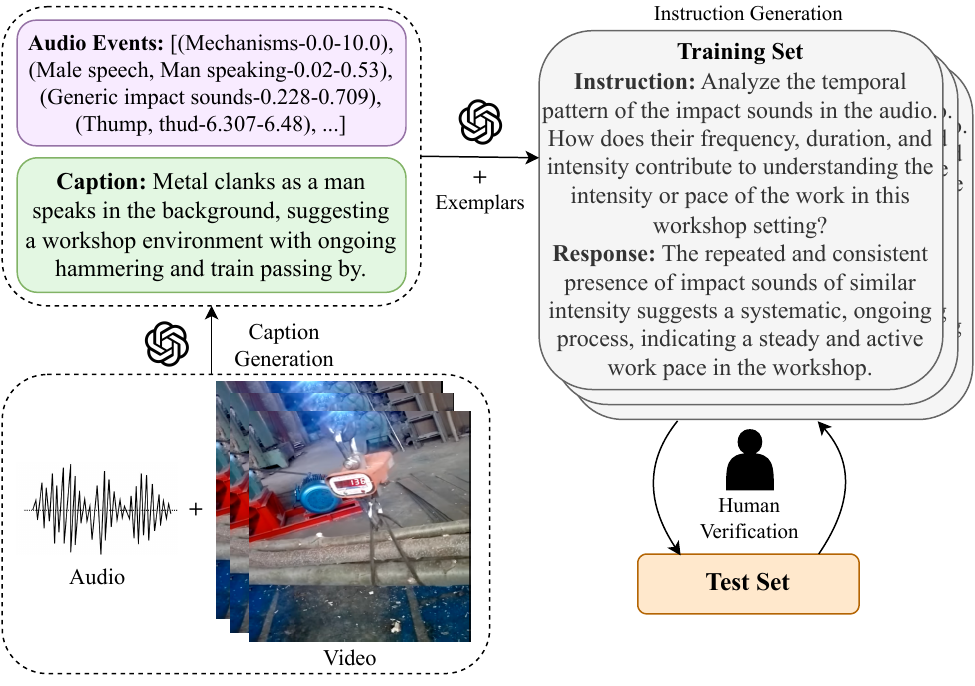}
    \caption{\small \textbf{Pipeline for synthesizing CompA-R.} For an audio in the AudioSet-strong dataset, we first use the audio and its corresponding video to generate a caption (described in Section~\ref{subsubsec:compar}). This caption is then fed into GPT-4 together with the ground-truth time slices for each event in the audio (available for AudioSet-strong). We then instruct GPT-4 to generate instruction-response pairs where the instruction is accompanied by human-written exemplars (prompt in Fig.~\ref{fig:llama_prompts}). To generate the test set, we perform expert human verification for a subset of the generated dataset.}
    \label{fig:gpt-instruc}
    \vspace{-3mm}
\end{figure}

\subsection{Training}

{\noindent \textbf{Fine-tuning.}} We fine-tune \texttt{GAMA} on the OpenAQA training set released by ~\citet{gong2024listen}. We use a faction of all the instances due to the unavailability of the entire AudioSet and resource constraints. Dataset details are provided in Appendix~\ref{sec:additional_details_general_appendix_gama_training}. Additionally, we augmented OpenAQA with 4 more datasets, including MusicCaps, MusicQA, NSynth, and Magna, to improve its music understanding capabilities. For fine-tuning, we follow the exact same 4-stage method proposed by~\citet{gong2024listen} where all parameters of all encoders are trainable, and we train only the LoRA modules of the LLM. We request our readers to refer to \citet{gong2024listen} for more details.

\begin{table*}[t]
\setlength\tabcolsep{1.1pt}
\centering
\fontsize{6.2}{6.2}\selectfont
\begin{tabular}{@{}lccccccccccccccccc@{}}
\toprule \midrule
\multicolumn{1}{c}{\textbf{Model}}& \begin{tabular}[c]{@{}c@{}}ESC50$^\#$ \\(Acc)\end{tabular} & \begin{tabular}[c]{@{}c@{}}DCASE$^\#$ \\ (Mi-F1)\end{tabular} & \begin{tabular}[c]{@{}c@{}}VS\textsuperscript{†} \\ (Acc)\end{tabular} & \begin{tabular}[c]{@{}c@{}}TUT\textsuperscript{†} \\ (Acc)\end{tabular} & \begin{tabular}[c]{@{}c@{}}BJO\textsuperscript{†} \\ (Acc)\end{tabular} & \multicolumn{1}{l}{\begin{tabular}[c]{@{}l@{}}VGG \\ (Acc)\end{tabular}} & {\begin{tabular}[c]{@{}c@{}}FSD \\ (mAP)\end{tabular}} & \begin{tabular}[c]{@{}c@{}}NS\textsubscript{ins.} \\(ACC)\end{tabular} & \begin{tabular}[c]{@{}c@{}}NS\textsubscript{src.} \\(ACC)\end{tabular} & \begin{tabular}[c]{@{}c@{}}GTZAN\textsuperscript{†} \\(ACC)\end{tabular} & \begin{tabular}[c]{@{}c@{}}MSD\textsuperscript{†} \\(ACC)\end{tabular} & { \begin{tabular}[c]{@{}c@{}}AudioSet\\ (mAP)\end{tabular}} & \begin{tabular}[c]{@{}c@{}}Classif.\\ Avg.\end{tabular} & \begin{tabular}[c]{@{}c@{}}AudioCaps \\(SPICE)\end{tabular} & \begin{tabular}[c]{@{}c@{}}Clotho \\(SPICE)\end{tabular} & \begin{tabular}[c]{@{}c@{}}Cap.\\ Avg.\end{tabular} & \begin{tabular}[c]{@{}c@{}}ClothoAQA \\(ACC)\end{tabular}\\\midrule
\multicolumn{14}{l}{\textit{Audio-Language encoder-based models. They are generalizable to unseen labels, but a pre-defined label set is required for inference.}}              \\
AudioCLIP               & 69.4                       & -    & -& -& -& -   { -}& - & -    & -        & -     & - & - & -        & -     & - & - \\
CLAP~\citep{elizalde2023clap}     & 82.6                       & 30.0& 48.4                    & 29.6                     & 47.5                     & 24.0                    & { 30.2} & 22.7 & 16.4 & 25.0  &   44.0           & { 5.8} & 29.4& -        & -     & - & - \\ 
CLAP~\cite{wu2023large}     & 89.1       & 31.3 & 47.1      & 35.6                    & 48.0                     & 26.3                    & 30.8 & 25.2 & 18.9 & 26.3 &    46.9        & 6.2  & 36.0 & -        & -     & - & - \\
CompA-CLAP                   & 90.1        & 30.6 & 49.5                    & 35.8                     & 48.2                     & 29.5                    & {31.5}             &24.9 & 17.0 & 26.1 &    46.2    & 6.2  & 36.3 & -        & -     & - & - \\ \midrule \midrule
\multicolumn{14}{l}{\textit{Audio-Language generation-based models. They directly output label names and do not need a pre-defined label set is needed at inference.}}  \\ 
\textcolor{black!50}{Qwen-Audio-Chat}  & \textcolor{black!50}{71.7}        & \textcolor{black!50}{32.4} & \textcolor{black!50}{74.2}          & \textcolor{black!50}{16.9}   & \textcolor{black!50}{50.8}     & \textcolor{black!50}{17.5}   & \textcolor{black!50}{39.8}  & \textcolor{black!50}{30.2}    & \textcolor{black!50}{41.3}  &  \textcolor{black!50}{41.6}   &  \textcolor{black!50}{69.1}  &  \textcolor{black!50}{13.4}   & \textcolor{black!50}{41.1} & \textcolor{black!50}{14.7} & \textcolor{black!50}{9.8} & \textcolor{black!50}{12.3}  & \textcolor{black!50}{32.3}\\
LTU  & \underline{81.7} & 37.5 & \textbf{53.3} & 19.9  & 67.8  & 50.3 & {43.9} & 28.0 & 41.8 & 9.9 & 74.2 & 18.3& 42.4 & 16.9   & 11.7   & \underline{15.8} & 25.1  \\
SALMONN   & 16.4\textsuperscript{†}  & 18.0\textsuperscript{†} & 16.9\textsuperscript{†}  & 7.8\textsuperscript{†}   & 25.0\textsuperscript{†}  & 23.3\textsuperscript{†}   & 22.1\textsuperscript{†} &   16.2\textsuperscript{†}  &  33.7\textsuperscript{†}  &  10.1\textsuperscript{†} & 28.8\textsuperscript{†}  &  13.4\textsuperscript{†}  &  17.9 & 8.3 &7.6 &8.0 &  23.1\textsuperscript{†}\\
Pengi  & 80.8\textsuperscript{†}& 29.6\textsuperscript{†} & 46.4\textsuperscript{†}  & 18.4\textsuperscript{†}  & 47.3\textsuperscript{†}  & 16.6\textsuperscript{†}     &   35.8     & 39.2 &  46.0  &  11.9    &   \textbf{93.0}    &   11.5    &   39.7    &12.7 & 7.0  & 9.9  &   63.6\\
AudioGPT  & 41.3  & 20.9 & 35.8  & 14.9  & 21.6 & 5.6 & 18.8 & 40.9 & 15.6 &  11.9  &  28.5    &  12.7  & 22.4 &6.9 & 6.2  & 6.6 & 33.4\\ 
\rowcolor[HTML]{f1e1f5} \texttt{GAMA}  \textit{(ours)} & \textbf{82.6}   & \textbf{38.4} & \underline{52.4} & \underline{21.5} & \textbf{69.5}  & \textbf{52.2} & \textbf{47.8}  & \textbf{63.9} & \textbf{99.5} & \textbf{13.8} &  \underline{85.6}   &  \textbf{19.2} & \textbf{53.9} & \textbf{18.5} & \textbf{13.5} & \textbf{16.0} &  \textbf{71.6}\\
\rowcolor[HTML]{f1e1f5} \hspace{2em} w/o AST \& Aggregator                  & 80.5   & 36.9 & 51.6 & 19.2 & 66.2  & 50.8 & 45.3 &  62.4  & 89.6 &  11.6  & {83.2} & 17.3 & 51.2 & 17.2 & 12.4 & 14.8 & 68.3\\
\rowcolor[HTML]{f1e1f5} \hspace{2em} w/ Last Layer Features  & 81.3   & \underline{37.6} & 50.2 & 20.4 & 68.2  & 51.7 &  45.8 &   \underline{62.6}  & \underline{92.3} & 11.2  & 81.5 & 18.1 & 51.7 & \underline{17.7} & 12.8 & 15.3 & \underline{69.5}\\
\rowcolor[HTML]{f1e1f5} \hspace{2em} w/o Audio Q-Former & 79.7   & 37.4 & 51.3 & 20.2 & 68.0  & 51.6   &   46.4   & 60.1  & 90.4  &  11.6  & 79.8 &  18.4      & 51.2 & 16.9 & 11.9 & 14.4 & 61.2 \\
\rowcolor[HTML]{f1e1f5} \hspace{2em} w/ CLAP  & 81.8   & 38.4 & 52.2 & \textbf{21.6} & \underline{69.1}  & \underline{52.0}   &    \underline{47.5}   & 58.8  & 99.5  &  \underline{12.4}  &  77.9    & \underline{19.0}      &   \underline{52.5}      & 17.2 & \underline{13.1} & 15.1 & 66.4\\
\bottomrule
\end{tabular}
\vspace{-0.75em}
\caption{\small Comparison of \texttt{GAMA} with baselines on evaluation datasets described on close-ended general audio and music understanding benchmarks. \texttt{GAMA} outperforms most ALMs on most settings. \textsuperscript{†} and $^\#$ indicate zero-shot and weak zero-shot, respectively. \textbf{Note:} Qwen-Audio-Chat does not provide training details. We also mark baseline values which are zero-shot.}
\label{tab:close_res}
\end{table*}

\begin{table*}[]
\centering
\resizebox{0.75\textwidth}{!}{
\begin{tabular}{l|cccc||cccc||ccc}
\toprule \hline
& \multicolumn{4}{c||}{\small\textbf{\texttt{CompA-R}-\textit{test}} (GPT-4/Human)} & \multicolumn{4}{c||}{\small\textbf{OpenAQA}} & \multicolumn{3}{c}{\small\textbf{Dense Captioning}}\\ 
\textbf{Models}     & Clarity & Correctness & Engagement & Avg. &Clarity & Correctness & Engagement & Avg. & AudioCaps & Clotho & Avg. \\ \hline
\textcolor{black!50}{Qwen-Audio-Chat} &     \textcolor{black!50}{3.5 / 3.4}       &    \textcolor{black!50}{3.3 / 3.4}    &      \textcolor{black!50}{\underline{3.6} / \underline{3.7}}   & \textcolor{black!50}{3.5 / 3.5}  & \textcolor{black!50}{3.6}     &     \textcolor{black!50}{3.6} & \textcolor{black!50}{3.5}   & \textcolor{black!50}{3.6} & \textcolor{black!50}{3.8}  & \textcolor{black!50}{3.6}  & \textcolor{black!50}{3.7}\\
LTU        &     3.5 / 4.0       &    3.2 / 3.3    &      3.4 / 3.5    &  3.4 / 3.6  &    3.5     &     3.7 &  3.5 & 3.6 &   3.5  &  3.6 & 3.5\\
SALMONN &      2.6 / 2.8       &     2.4 / 2.3    &      2.0 / 2.2    &  2.3 / 2.4   &   2.4      &   2.5 & 2.7  & 2.5 &   2.8   &  3.1 & 2.9\\
Pengi      &      1.8 / 1.6      &     1.5 / 1.4   &      1.3 / 1.2   &  1.5 / 1.4   &   1.7      &   1.5   &   1.4  & 1.5 &  2.6  &  2.8    & 2.7\\
AudioGPT   &     1.3  / 1.4      &     1.6 / 1.5   &      1.4 / 1.7   &  1.4 / 1.5  &   1.6      &    1.5     & 1.5    & 1.5 & 2.7 & 2.9 & 2.8\\
LTU w/ \texttt{CompA-R}      &     3.5 / 4.0       &    3.2 / 3.3    &      3.4 / 3.5        & 3.6 /3.6  &  3.5     &     3.7 &  3.5 & 3.6 &   3.7  &  3.8 & 3.8\\
\rowcolor[HTML]{f1e1f5}\texttt{GAMA-IT} \textit{(ours)}     &      \textbf{4.3 / 4.5} & \textbf{3.9 / 4.1}     &    \textbf{3.9 / 4.3}  & \textbf{4.0 / 4.3}  &   \textbf{4.0}    &     \textbf{4.2}    &      \textbf{3.8}  & \textbf{4.0} &   \textbf{4.3}  & \textbf{4.1}& \textbf{4.2}\\ 
\rowcolor[HTML]{f1e1f5} \hspace{2em} w/o Soft Prompt  &      \underline{4.1} / \underline{4.2} &  \underline{3.7} / \underline{3.8}     &    \underline{3.6} / 3.4    & \underline{3.8} / \underline{3.8} & \underline{3.9}    &     \underline{3.8}    &      \underline{3.7}   &  \underline{3.8}  & \underline{4.1}  & \underline{3.9} & \underline{4.0}\\
\rowcolor[HTML]{f1e1f5} \hspace{2em} w/o Aggregator   &      4.0 / 4.2 &  3.5 / 3.5     &    3.6 / 3.5  & 3.7 / 3.7 &   3.7    &     3.7    &   3.5 & 3.6 &  3.7  & 3.8 & 3.8  \\
\rowcolor[HTML]{f1e1f5} \hspace{2em} w/o Audio Q-Former    &    3.8 / 3.7 &  3.4 / 3.6     &    3.5 / 3.3    & 3.6 / 3.5 & 3.4    &  3.9  &  3.5 & 3.6 &  3.7  & 3.5 &  3.6  \\
\hline
\end{tabular}}
\vspace{-0.75em}
\caption{\small Comparison of \texttt{GAMA} with other baselines on open-ended AQA on OpenAQA, complex open-ended AQA on \texttt{CompA-R}-\textit{test} and Dense Captioning on 500 instances from AudioCaps and Clotho.}
\label{tab:aqa_res}
\end{table*}

{\noindent \textbf{Instruction Tuning on \texttt{CompA-R}.}} Post fine-tuning, we instruction-tune \texttt{GAMA} on \texttt{CompA-R} to endow it with complex reasoning abilities. Following common conventions~\cite{liu2023visual}, we fine-tuned only the LoRA modules. We call the Instruction Tuned \texttt{GAMA} as \texttt{GAMA-IT}. Although fine-tuning on AQA also endows \texttt{GAMA} with instruction-following capabilities, \texttt{CompA-R} differs in the nature of training instances (thereby the capabilities it endows), and thus, we differentiate with such a naming convention for ease of reading.

\subsection{Experimental Setup}
{\noindent \textbf{Hyper-parameters.}} For the fine-tuning stage, we follow the exact same hyper-parameter setup proposed by~\citet{gong2024listen}. However, we scale down our batch sizes to 4, 2, 2, and 2 (due to compute constraints) with an effective batch size of 256 in all stages. For Instruction Tuning, we employ a batch size of 2, an effective batch size of 256, and a learning rate of 1e-4. For both training and evaluation, we sampled audio at 16kHz.
\vspace{0.5mm}

{\noindent \textbf{Baselines.}} We compare \texttt{GAMA} with \textit{\textbf{i)}} generation-based LALMs: LTU, Qwen-Audio, SALMONN, Pengi and AudioGPT. We only employ the original checkpoints open-sourced by the authors and do not re-train the models due to compute constraints (except LTU, which we retrain on our version of OpenAQA, the same batch size as GAMA, and with LLaMa-2 as the LLM). We do not compare with Audio Flamingo~\cite{kong2024audio} as the checkpoint was not available at the time of writing the paper, and we are constrained by compute for training it from scratch. \textit{\textbf{ii)}} audio-language encoders: CLAP by ~\citet{laionclap2023} and ~\citet{10095889}, CompA-CLAP~\cite{ghosh2024compa}, AudioCLIP~\cite{guzhov2021audioclip} and Audio Q-Former. For dense captioning and close- and open-ended AQA, we evaluate using \texttt{GAMA-IT}. For all other tasks, we evaluate using the only fine-tuned version of \texttt{GAMA} (rationale in Appendix~\ref{sec:gamait}).
\vspace{0.5mm}

{\noindent \textbf{Evaluation Datasets and Metrics.}} Evaluation metrics used for all evaluation datasets are mentioned in Table~\ref{tab:aqa_res} and detailed statistics about each dataset is mentioned in Section~\ref{sec:gama_eval_additional_details}. For classification, zero-shot evaluation refers to datasets \texttt{GAMA} that have never been seen during training; weak zero-shot evaluation refers to datasets \texttt{GAMA} that have not been seen in training but are sourced from the same project as part of the training data, and seen datasets refer to datasets \texttt{GAMA} has been trained on. Similar to ~\citet{deshmukh2023pengi,gong2024listen}, we first caption the audio and retrieve the most similar label using SentenceBERT. We employ either accuracy (Acc), Micro-F1 (Mi-F1), or Mean Average Precision (mAP) for classification evaluation. For captioning, we also propose dense captioning, which evaluates a model for its capability to identify every event in the audio and the context of its occurrence with respect to other events in the audio (more in Section~\ref{sec:results}). For evaluation, we randomly select a subset of 500 samples from AudioCaps and Clotho. We also employ human evaluation for OpenAQA, \texttt{CompA-R}-\textit{test}, and dense captioning. For human evaluation, we ask human annotators to score the caption on a scale of 1-5 and report the score averaged across the 3. More details on recruitment and background of annotators can be found in Appendix~\ref{sec:human_study}. Finally, due to human evaluation being prohibitively expensive, we also propose an automated evaluation methodology for complex open-ended AQA on \texttt{CompA-R}-\textit{test}. We evaluate model responses using text-only GPT-4, where we provide it with the audio caption generated in Section~\ref{subsubsec:compar} and the gold-standard audio event with timestamps (prompt in Appendix~\ref{sec:prompts}).

\section{Results and Analysis}
\label{sec:results}

{\noindent \textbf{Quantitative Results.}} Table~\ref{tab:close_res} compares \texttt{GAMA} with other baselines on classification and captioning tasks. For zero-shot classification evaluation on VocalSound (VS)~\cite{gong2022vocalsound}, TUT 2017 (TUT)~\cite{mesaros2018multi}, Beijing Opera (BJO)~\cite{tian2014study}, GTZAN (GTZ)~\cite{park2022enemy} and Medley-solos-DB (MDB)~\cite{lostanlen2018medley}, \texttt{GAMA} outperforms our baselines by 2\%-67\%. For weak zero-shot evaluation on ESC-50~\cite{piczak2015esc} and DCASE2017 Task 4 (DCASE)~\cite{mesaros2017dcase}, \texttt{GAMA} outperforms our baselines by 1\%-66\%. Finally, for in-domain evaluation on VGGSound (VGG)~\cite{chen2020vggsound}, FSD50K (FSD)~\cite{fonseca2021fsd50k}, AudioSet (AS)~\cite{gemmeke2017audio} and NSynth (NS)~\cite{engel2017neural} \texttt{GAMA} outperforms our baselines by 1\%-84\%. \texttt{GAMA} sees the steepest drop in performance when the AST and Aggregator are removed (i.e., only Auio Q-Former is employed).

Table~\ref{tab:aqa_res} compares \texttt{GAMA} with other baselines on AQA (open-ended and complex open-ended) and dense captioning. \texttt{GAMA} outperforms all our baselines on all settings. \texttt{GAMA} shows absolute improvement of 4\% - 50\% on OpenAQA, 8\% - 58\% on \texttt{CompA-R}-\textit{test} and 8\% - 30\% on Dense Captioning. Similar to the tasks in Table~\ref{tab:close_res}, performance on benchmarks suffers the most when without the Audio Q-Former (when only the AST and Aggregator are employed). Audio Q-Former proves to especially effective (over employing CLAP) in AQA.

{\noindent \textbf{Qualitative Results.}} Fig.~\ref{fig:qa_example1} compares \texttt{GAMA-IT} against other LALMs from literature with instances from \texttt{CompA-R}-\textit{test}. All models compared by default possess audio chat or open-ended AQA capabilities.  \texttt{GAMA-IT} is able to provide more faithful responses that are both correct and preferred more by humans. We provide additional comparisons in Figs.~\ref{fig:app1}, ~\ref{fig:app2}, ~\ref{fig:app3}, ~\ref{fig:app4}, ~\ref{fig:app5}, and our demo page:  (where we also show comparisons of dense captioning).

\section{Conclusion}
In this paper, we propose \texttt{GAMA}, an LALM with improved audio perception abilities. We integrate an LLM with multiple types of audio representations, which are responsible for providing diverse knowledge about the input audio. \texttt{GAMA} fine-tuned on a mixture of open-source datasets outperforms prior audio-language models by significant margins on 16 datasets spanning 4 tasks. Next, we propose \texttt{CompA-R}, an instruction-tuning dataset that we synthesize using a robust pipeline for endowing an LALM with complex reasoning abilities. \texttt{GAMA} IT-ed on \texttt{CompA-R} outperforms baselines on complex open-ended AQA and dense captioning.

\section*{Limitations and Future Work}

\texttt{GAMA} and our experimental setup have several limitations, including:
\begin{itemize}
    \item For the scope of our experiments, we do not evaluate and compare music understanding extensively. We do not do this as we do not train \texttt{GAMA} on diverse and large-scale music datasets. We also acknowledge that it is possible to employ the \texttt{GAMA} architecture for comprehensive music understanding if trained on large-scale music understanding datasets. As part of future work, we plan to release a music-only version of \texttt{GAMA}, similar to ~\citet{gardner2024llark}.

    \item We do not employ larger LLMs, for example, the 13B versions of the LLaMA family, similar to ~\citet{tang2024salmonn} and ~\citet{gong2024listen}, due to compute constraints.

    \item The audio-encoder(s) in \texttt{GAMA} have more parameters than in our baselines. However, we also acknowledge that this adds to only a fraction of the total parameter count of the LALM.

\end{itemize}

\bibliography{anthology,custom}

\begin{thebibliography}{62}
\expandafter\ifx\csname natexlab\endcsname\relax\def\natexlab#1{#1}\fi

\bibitem[{BBC(2018)}]{BBCSoundEffects2018}
 2018.
\newblock \href {http://bbcsfx.acropolis.org.uk/} {{A dump of BBC's sound effects library}}.
\newblock This dump was created using the script found at https://github.com/FThompson/BBCSoundDownloader. Identifier: BBCSoundEffectsComplete.

\bibitem[{sou(2023)}]{soundbible}
 2023.
\newblock \href {https://soundbible.com/} {{SoundBible - Free Sound Clips, Sound Bites, and Sound Effects}}.
\newblock Accessed: 25 September 2023.

\bibitem[{Agostinelli et~al.(2023)Agostinelli, Denk, Borsos, Engel, Verzetti, Caillon, Huang, Jansen, Roberts, Tagliasacchi, Sharifi, Zeghidour, and Frank}]{agostinelli2023musiclm}
Andrea Agostinelli, Timo~I. Denk, Zalán Borsos, Jesse Engel, Mauro Verzetti, Antoine Caillon, Qingqing Huang, Aren Jansen, Adam Roberts, Marco Tagliasacchi, Matt Sharifi, Neil Zeghidour, and Christian Frank. 2023.
\newblock \href {http://arxiv.org/abs/2301.11325} {Musiclm: Generating music from text}.

\bibitem[{Bubeck et~al.(2023)Bubeck, Chandrasekaran, Eldan, Gehrke, Horvitz, Kamar, Lee, Lee, Li, Lundberg et~al.}]{bubeck2023sparks}
S{\'e}bastien Bubeck, Varun Chandrasekaran, Ronen Eldan, Johannes Gehrke, Eric Horvitz, Ece Kamar, Peter Lee, Yin~Tat Lee, Yuanzhi Li, Scott Lundberg, et~al. 2023.
\newblock Sparks of artificial general intelligence: Early experiments with gpt-4.
\newblock \emph{arXiv preprint arXiv:2303.12712}.

\bibitem[{Chen et~al.(2020)Chen, Xie, Vedaldi, and Zisserman}]{chen2020vggsound}
Honglie Chen, Weidi Xie, Andrea Vedaldi, and Andrew Zisserman. 2020.
\newblock \href {http://arxiv.org/abs/2004.14368} {Vggsound: A large-scale audio-visual dataset}.

\bibitem[{Chu et~al.(2023)Chu, Xu, Zhou, Yang, Zhang, Yan, Zhou, and Zhou}]{Qwen-Audio}
Yunfei Chu, Jin Xu, Xiaohuan Zhou, Qian Yang, Shiliang Zhang, Zhijie Yan, Chang Zhou, and Jingren Zhou. 2023.
\newblock Qwen-audio: Advancing universal audio understanding via unified large-scale audio-language models.
\newblock \emph{arXiv preprint arXiv:2311.07919}.

\bibitem[{Cui and Wang(2024)}]{cui2024adainstruct}
Wanyun Cui and Qianle Wang. 2024.
\newblock \href {https://openreview.net/forum?id=O04DqGdAqQ} {Ada-instruct: Adapting instruction generators for complex reasoning}.

\bibitem[{Deng et~al.(2009)Deng, Dong, Socher, Li, Li, and Fei-Fei}]{deng2009imagenet}
Jia Deng, Wei Dong, Richard Socher, Li-Jia Li, Kai Li, and Li~Fei-Fei. 2009.
\newblock Imagenet: A large-scale hierarchical image database.
\newblock In \emph{2009 IEEE conference on computer vision and pattern recognition}, pages 248--255. Ieee.

\bibitem[{Deshmukh et~al.(2023)Deshmukh, Elizalde, Singh, and Wang}]{deshmukh2023pengi}
Soham Deshmukh, Benjamin Elizalde, Rita Singh, and Huaming Wang. 2023.
\newblock \href {http://arxiv.org/abs/2305.11834} {Pengi: An audio language model for audio tasks}.

\bibitem[{Deshmukh et~al.(2022)Deshmukh, Elizalde, and Wang}]{deshmukh2022audio}
Soham Deshmukh, Benjamin Elizalde, and Huaming Wang. 2022.
\newblock Audio retrieval with wavtext5k and clap training.
\newblock \emph{arXiv preprint arXiv:2209.14275}.

\bibitem[{Devlin et~al.(2018)Devlin, Chang, Lee, and Toutanova}]{devlin2018bert}
Jacob Devlin, Ming-Wei Chang, Kenton Lee, and Kristina Toutanova. 2018.
\newblock Bert: Pre-training of deep bidirectional transformers for language understanding.
\newblock \emph{arXiv preprint arXiv:1810.04805}.

\bibitem[{Drossos et~al.(2020)Drossos, Lipping, and Virtanen}]{drossos2020clotho}
Konstantinos Drossos, Samuel Lipping, and Tuomas Virtanen. 2020.
\newblock Clotho: An audio captioning dataset.
\newblock In \emph{ICASSP 2020-2020 IEEE International Conference on Acoustics, Speech and Signal Processing (ICASSP)}, pages 736--740. IEEE.

\bibitem[{Elizalde et~al.(2023{\natexlab{a}})Elizalde, Deshmukh, Al~Ismail, and Wang}]{elizalde2023clap}
Benjamin Elizalde, Soham Deshmukh, Mahmoud Al~Ismail, and Huaming Wang. 2023{\natexlab{a}}.
\newblock Clap learning audio concepts from natural language supervision.
\newblock In \emph{ICASSP 2023-2023 IEEE International Conference on Acoustics, Speech and Signal Processing (ICASSP)}, pages 1--5. IEEE.

\bibitem[{Elizalde et~al.(2023{\natexlab{b}})Elizalde, Deshmukh, Ismail, and Wang}]{10095889}
Benjamin Elizalde, Soham Deshmukh, Mahmoud~Al Ismail, and Huaming Wang. 2023{\natexlab{b}}.
\newblock \href {https://doi.org/10.1109/ICASSP49357.2023.10095889} {Clap learning audio concepts from natural language supervision}.
\newblock In \emph{ICASSP 2023 - 2023 IEEE International Conference on Acoustics, Speech and Signal Processing (ICASSP)}, pages 1--5.

\bibitem[{Engel et~al.(2017)Engel, Resnick, Roberts, Dieleman, Norouzi, Eck, and Simonyan}]{engel2017neural}
Jesse Engel, Cinjon Resnick, Adam Roberts, Sander Dieleman, Mohammad Norouzi, Douglas Eck, and Karen Simonyan. 2017.
\newblock Neural audio synthesis of musical notes with wavenet autoencoders.
\newblock In \emph{International Conference on Machine Learning}, pages 1068--1077. PMLR.

\bibitem[{Fonseca et~al.(2021)Fonseca, Favory, Pons, Font, and Serra}]{fonseca2021fsd50k}
Eduardo Fonseca, Xavier Favory, Jordi Pons, Frederic Font, and Xavier Serra. 2021.
\newblock Fsd50k: an open dataset of human-labeled sound events.
\newblock \emph{IEEE/ACM Transactions on Audio, Speech, and Language Processing}, 30:829--852.

\bibitem[{Fonseca et~al.(2022)Fonseca, Favory, Pons, Font, and Serra}]{fonseca2022fsd50k}
Eduardo Fonseca, Xavier Favory, Jordi Pons, Frederic Font, and Xavier Serra. 2022.
\newblock \href {http://arxiv.org/abs/2010.00475} {Fsd50k: An open dataset of human-labeled sound events}.

\bibitem[{Gardner et~al.(2024)Gardner, Durand, Stoller, and Bittner}]{gardner2024llark}
Joshua~P Gardner, Simon Durand, Daniel Stoller, and Rachel~M Bittner. 2024.
\newblock \href {https://openreview.net/forum?id=7fwzPsn1lJ} {{LL}ark: A multimodal foundation model for music}.

\bibitem[{Gemmeke et~al.(2017)Gemmeke, Ellis, Freedman, Jansen, Lawrence, Moore, Plakal, and Ritter}]{gemmeke2017audio}
Jort~F Gemmeke, Daniel~PW Ellis, Dylan Freedman, Aren Jansen, Wade Lawrence, R~Channing Moore, Manoj Plakal, and Marvin Ritter. 2017.
\newblock Audio set: An ontology and human-labeled dataset for audio events.
\newblock In \emph{2017 IEEE international conference on acoustics, speech and signal processing (ICASSP)}, pages 776--780. IEEE.

\bibitem[{Ghosh et~al.(2024{\natexlab{a}})Ghosh, Kumar, Reddy~Evuru, Duraiswami, and Manocha}]{10448030}
Sreyan Ghosh, Sonal Kumar, Chandra~Kiran Reddy~Evuru, Ramani Duraiswami, and Dinesh Manocha. 2024{\natexlab{a}}.
\newblock \href {https://doi.org/10.1109/ICASSP48485.2024.10448030} {Recap: Retrieval-augmented audio captioning}.
\newblock In \emph{ICASSP 2024 - 2024 IEEE International Conference on Acoustics, Speech and Signal Processing (ICASSP)}, pages 1161--1165.

\bibitem[{Ghosh et~al.(2024{\natexlab{b}})Ghosh, Seth, Kumar, Tyagi, Evuru, S, Sakshi, Nieto, Duraiswami, and Manocha}]{ghosh2024compa}
Sreyan Ghosh, Ashish Seth, Sonal Kumar, Utkarsh Tyagi, Chandra Kiran~Reddy Evuru, Ramaneswaran S, S~Sakshi, Oriol Nieto, Ramani Duraiswami, and Dinesh Manocha. 2024{\natexlab{b}}.
\newblock \href {https://openreview.net/forum?id=86NGO8qeWs} {Compa: Addressing the gap in compositional reasoning in audio-language models}.
\newblock In \emph{The Twelfth International Conference on Learning Representations}.

\bibitem[{Gong et~al.(2021)Gong, Chung, and Glass}]{gong2021ast}
Yuan Gong, Yu-An Chung, and James Glass. 2021.
\newblock Ast: Audio spectrogram transformer.
\newblock \emph{arXiv preprint arXiv:2104.01778}.

\bibitem[{Gong et~al.(2024)Gong, Luo, Liu, Karlinsky, and Glass}]{gong2024listen}
Yuan Gong, Hongyin Luo, Alexander~H. Liu, Leonid Karlinsky, and James~R. Glass. 2024.
\newblock \href {https://openreview.net/forum?id=nBZBPXdJlC} {Listen, think, and understand}.
\newblock In \emph{The Twelfth International Conference on Learning Representations}.

\bibitem[{Gong et~al.(2023)Gong, Rouditchenko, Liu, Harwath, Karlinsky, Kuehne, and Glass}]{gong2023contrastive}
Yuan Gong, Andrew Rouditchenko, Alexander~H. Liu, David Harwath, Leonid Karlinsky, Hilde Kuehne, and James~R. Glass. 2023.
\newblock \href {https://openreview.net/forum?id=QPtMRyk5rb} {Contrastive audio-visual masked autoencoder}.
\newblock In \emph{The Eleventh International Conference on Learning Representations}.

\bibitem[{Gong et~al.(2022)Gong, Yu, and Glass}]{gong2022vocalsound}
Yuan Gong, Jin Yu, and James Glass. 2022.
\newblock Vocalsound: A dataset for improving human vocal sounds recognition.
\newblock In \emph{ICASSP 2022-2022 IEEE International Conference on Acoustics, Speech and Signal Processing (ICASSP)}, pages 151--155. IEEE.

\bibitem[{Gudibande et~al.(2023)Gudibande, Wallace, Snell, Geng, Liu, Abbeel, Levine, and Song}]{gudibande2023false}
Arnav Gudibande, Eric Wallace, Charlie Snell, Xinyang Geng, Hao Liu, Pieter Abbeel, Sergey Levine, and Dawn Song. 2023.
\newblock The false promise of imitating proprietary llms.
\newblock \emph{arXiv preprint arXiv:2305.15717}.

\bibitem[{Guzhov et~al.(2022)Guzhov, Raue, Hees, and Dengel}]{audioclip}
Andrey Guzhov, Federico Raue, J{\"{o}}rn Hees, and Andreas Dengel. 2022.
\newblock Audioclip: Extending clip to image, text and audio.
\newblock In \emph{{ICASSP} 2022}.

\bibitem[{Guzhov et~al.(2021)Guzhov, Raue, Hees, and Dengel}]{guzhov2021audioclip}
Andrey Guzhov, Federico Raue, Jörn Hees, and Andreas Dengel. 2021.
\newblock \href {http://arxiv.org/abs/2106.13043} {Audioclip: Extending clip to image, text and audio}.

\bibitem[{Hu et~al.(2024)Hu, CHEN, Yang, Li, Zhang, Chen, and Chng}]{hu2024large}
Yuchen Hu, CHEN CHEN, Chao-Han~Huck Yang, Ruizhe Li, Chao Zhang, Pin-Yu Chen, and Ensiong Chng. 2024.
\newblock \href {https://openreview.net/forum?id=ceATjGPTUD} {Large language models are efficient learners of noise-robust speech recognition}.
\newblock In \emph{The Twelfth International Conference on Learning Representations}.

\bibitem[{Huang et~al.(2024)Huang, Li, Yang, Shi, Chang, Ye, Wu, Hong, Huang, Liu et~al.}]{huang2024audiogpt}
Rongjie Huang, Mingze Li, Dongchao Yang, Jiatong Shi, Xuankai Chang, Zhenhui Ye, Yuning Wu, Zhiqing Hong, Jiawei Huang, Jinglin Liu, et~al. 2024.
\newblock Audiogpt: Understanding and generating speech, music, sound, and talking head.
\newblock In \emph{Proceedings of the AAAI Conference on Artificial Intelligence}, volume~38, pages 23802--23804.

\bibitem[{Kim et~al.(2019)Kim, Kim, Lee, and Kim}]{kim2019audiocaps}
Chris~Dongjoo Kim, Byeongchang Kim, Hyunmin Lee, and Gunhee Kim. 2019.
\newblock Audiocaps: Generating captions for audios in the wild.
\newblock In \emph{Proceedings of the 2019 Conference of the North American Chapter of the Association for Computational Linguistics: Human Language Technologies, Volume 1 (Long and Short Papers)}, pages 119--132.

\bibitem[{Kojima et~al.(2022)Kojima, Gu, Reid, Matsuo, and Iwasawa}]{kojima2022large}
Takeshi Kojima, Shixiang~Shane Gu, Machel Reid, Yutaka Matsuo, and Yusuke Iwasawa. 2022.
\newblock Large language models are zero-shot reasoners.
\newblock \emph{Advances in neural information processing systems}, 35:22199--22213.

\bibitem[{Kong et~al.(2024)Kong, Goel, Badlani, Ping, Valle, and Catanzaro}]{kong2024audio}
Zhifeng Kong, Arushi Goel, Rohan Badlani, Wei Ping, Rafael Valle, and Bryan Catanzaro. 2024.
\newblock \href {http://arxiv.org/abs/2402.01831} {Audio flamingo: A novel audio language model with few-shot learning and dialogue abilities}.

\bibitem[{Li et~al.(2023)Li, Li, Savarese, and Hoi}]{li2023blip}
Junnan Li, Dongxu Li, Silvio Savarese, and Steven Hoi. 2023.
\newblock Blip-2: Bootstrapping language-image pre-training with frozen image encoders and large language models.
\newblock In \emph{International conference on machine learning}, pages 19730--19742. PMLR.

\bibitem[{Lipping et~al.(2022)Lipping, Sudarsanam, Drossos, and Virtanen}]{lipping2022clotho}
Samuel Lipping, Parthasaarathy Sudarsanam, Konstantinos Drossos, and Tuomas Virtanen. 2022.
\newblock Clotho-aqa: A crowdsourced dataset for audio question answering.
\newblock In \emph{2022 30th European Signal Processing Conference (EUSIPCO)}, pages 1140--1144. IEEE.

\bibitem[{Liu et~al.(2023{\natexlab{a}})Liu, Li, Li, and Lee}]{liu2023improved}
Haotian Liu, Chunyuan Li, Yuheng Li, and Yong~Jae Lee. 2023{\natexlab{a}}.
\newblock Improved baselines with visual instruction tuning.
\newblock \emph{arXiv preprint arXiv:2310.03744}.

\bibitem[{Liu et~al.(2023{\natexlab{b}})Liu, Li, Wu, and Lee}]{liu2023visual}
Haotian Liu, Chunyuan Li, Qingyang Wu, and Yong~Jae Lee. 2023{\natexlab{b}}.
\newblock \href {http://arxiv.org/abs/2304.08485} {Visual instruction tuning}.

\bibitem[{Liu et~al.(2024)Liu, Li, Wu, and Lee}]{liu2024visual}
Haotian Liu, Chunyuan Li, Qingyang Wu, and Yong~Jae Lee. 2024.
\newblock Visual instruction tuning.
\newblock \emph{Advances in neural information processing systems}, 36.

\bibitem[{Liu et~al.(2023{\natexlab{c}})Liu, Zeng, Ren, Li, Zhang, Yang, Li, Yang, Su, Zhu et~al.}]{liu2023grounding}
Shilong Liu, Zhaoyang Zeng, Tianhe Ren, Feng Li, Hao Zhang, Jie Yang, Chunyuan Li, Jianwei Yang, Hang Su, Jun Zhu, et~al. 2023{\natexlab{c}}.
\newblock Grounding dino: Marrying dino with grounded pre-training for open-set object detection.
\newblock \emph{arXiv preprint arXiv:2303.05499}.

\bibitem[{Lostanlen et~al.(2018)Lostanlen, Cella, Bittner, and Essid}]{lostanlen2018medley}
Vincent Lostanlen, Carmine-Emanuele Cella, Rachel Bittner, and Slim Essid. 2018.
\newblock Medley-solos-db: a crosscollection dataset for musical instrument recognition.
\newblock \emph{Zenodo}.

\bibitem[{Lostanlen et~al.(2019)Lostanlen, Cella, Bittner, and Essid}]{lostanlenvincent20193464194}
Vincent Lostanlen, Carmine-Emanuele Cella, Rachel Bittner, and Slim Essid. 2019.
\newblock \href {https://doi.org/10.5281/zenodo.3464194} {{Medley-solos-DB: a cross-collection dataset for musical instrument recognition}}.

\bibitem[{Mesaros et~al.(2017)Mesaros, Heittola, Diment, Elizalde, Shah, Vincent, Raj, and Virtanen}]{mesaros2017dcase}
Annamaria Mesaros, Toni Heittola, Aleksandr Diment, Benjamin Elizalde, Ankit Shah, Emmanuel Vincent, Bhiksha Raj, and Tuomas Virtanen. 2017.
\newblock Dcase 2017 challenge setup: Tasks, datasets and baseline system.
\newblock In \emph{DCASE 2017-Workshop on Detection and Classification of Acoustic Scenes and Events}.

\bibitem[{Mesaros et~al.(2018)Mesaros, Heittola, and Virtanen}]{mesaros2018multi}
Annamaria Mesaros, Toni Heittola, and Tuomas Virtanen. 2018.
\newblock A multi-device dataset for urban acoustic scene classification.
\newblock \emph{arXiv preprint arXiv:1807.09840}.

\bibitem[{Morato and Mesaros(2021)}]{irene_martin_morato_2021_5114771}
Irene~Martin Morato and Annamaria Mesaros. 2021.
\newblock \href {https://doi.org/10.5281/zenodo.5114771} {Macs - multi-annotator captioned soundscapes}.

\bibitem[{Park et~al.(2022)Park, Cho, Sim, Lee, and Choo}]{park2022enemy}
Junwoo Park, Youngwoo Cho, Gyuhyeon Sim, Hojoon Lee, and Jaegul Choo. 2022.
\newblock Enemy spotted: in-game gun sound dataset for gunshot classification and localization.
\newblock In \emph{2022 IEEE Conference on Games (CoG)}, pages 56--63. IEEE.

\bibitem[{Piczak(2015)}]{piczak2015esc}
Karol~J Piczak. 2015.
\newblock Esc: Dataset for environmental sound classification.
\newblock In \emph{Proceedings of the 23rd ACM international conference on Multimedia}, pages 1015--1018.

\bibitem[{Radford et~al.(2021)Radford, Kim, Hallacy, Ramesh, Goh, Agarwal, Sastry, Askell, Mishkin, Clark et~al.}]{radford2021learning}
Alec Radford, Jong~Wook Kim, Chris Hallacy, Aditya Ramesh, Gabriel Goh, Sandhini Agarwal, Girish Sastry, Amanda Askell, Pamela Mishkin, Jack Clark, et~al. 2021.
\newblock Learning transferable visual models from natural language supervision.
\newblock In \emph{International conference on machine learning}, pages 8748--8763. PMLR.

\bibitem[{Singla et~al.(2022)Singla, Shah, Chen, and Shah}]{singla2022audio}
Yaman~Kumar Singla, Jui Shah, Changyou Chen, and Rajiv~Ratn Shah. 2022.
\newblock What do audio transformers hear? probing their representations for language delivery \& structure.
\newblock In \emph{2022 IEEE International Conference on Data Mining Workshops (ICDMW)}, pages 910--925. IEEE.

\bibitem[{{Sonniss Limited}(2022)}]{sonniss2022}
{Sonniss Limited}. 2022.
\newblock \href {https://sonniss.com/gameaudiogdc} {{Sonniss Game Audio}}.
\newblock Registered in England, UK. Company number: 09377364. Accessed: 25 September 2023.

\bibitem[{Tang et~al.(2024)Tang, Yu, Sun, Chen, Tan, Li, Lu, MA, and Zhang}]{tang2024salmonn}
Changli Tang, Wenyi Yu, Guangzhi Sun, Xianzhao Chen, Tian Tan, Wei Li, Lu~Lu, Zejun MA, and Chao Zhang. 2024.
\newblock \href {https://openreview.net/forum?id=14rn7HpKVk} {{SALMONN}: Towards generic hearing abilities for large language models}.
\newblock In \emph{The Twelfth International Conference on Learning Representations}.

\bibitem[{Tian et~al.(2014)Tian, Srinivasamurthy, Sandler, and Serra}]{tian2014study}
Mi~Tian, Ajay Srinivasamurthy, Mark Sandler, and Xavier Serra. 2014.
\newblock A study of instrument-wise onset detection in beijing opera percussion ensembles.
\newblock In \emph{2014 ieee international conference on acoustics, speech and signal processing (icassp)}, pages 2159--2163. IEEE.

\bibitem[{Touvron et~al.(2023)Touvron, Martin, Stone, Albert, Almahairi, Babaei, Bashlykov, Batra, Bhargava, Bhosale, Bikel, Blecher, Ferrer, Chen, Cucurull, Esiobu, Fernandes, Fu, Fu, Fuller, Gao, Goswami, Goyal, Hartshorn, Hosseini, Hou, Inan, Kardas, Kerkez, Khabsa, Kloumann, Korenev, Koura, Lachaux, Lavril, Lee, Liskovich, Lu, Mao, Martinet, Mihaylov, Mishra, Molybog, Nie, Poulton, Reizenstein, Rungta, Saladi, Schelten, Silva, Smith, Subramanian, Tan, Tang, Taylor, Williams, Kuan, Xu, Yan, Zarov, Zhang, Fan, Kambadur, Narang, Rodriguez, Stojnic, Edunov, and Scialom}]{touvron2023llama}
Hugo Touvron, Louis Martin, Kevin Stone, Peter Albert, Amjad Almahairi, Yasmine Babaei, Nikolay Bashlykov, Soumya Batra, Prajjwal Bhargava, Shruti Bhosale, Dan Bikel, Lukas Blecher, Cristian~Canton Ferrer, Moya Chen, Guillem Cucurull, David Esiobu, Jude Fernandes, Jeremy Fu, Wenyin Fu, Brian Fuller, Cynthia Gao, Vedanuj Goswami, Naman Goyal, Anthony Hartshorn, Saghar Hosseini, Rui Hou, Hakan Inan, Marcin Kardas, Viktor Kerkez, Madian Khabsa, Isabel Kloumann, Artem Korenev, Punit~Singh Koura, Marie-Anne Lachaux, Thibaut Lavril, Jenya Lee, Diana Liskovich, Yinghai Lu, Yuning Mao, Xavier Martinet, Todor Mihaylov, Pushkar Mishra, Igor Molybog, Yixin Nie, Andrew Poulton, Jeremy Reizenstein, Rashi Rungta, Kalyan Saladi, Alan Schelten, Ruan Silva, Eric~Michael Smith, Ranjan Subramanian, Xiaoqing~Ellen Tan, Binh Tang, Ross Taylor, Adina Williams, Jian~Xiang Kuan, Puxin Xu, Zheng Yan, Iliyan Zarov, Yuchen Zhang, Angela Fan, Melanie Kambadur, Sharan Narang, Aurelien Rodriguez, Robert Stojnic, Sergey Edunov, and Thomas
  Scialom. 2023.
\newblock \href {http://arxiv.org/abs/2307.09288} {Llama 2: Open foundation and fine-tuned chat models}.

\bibitem[{Tzanetakis et~al.(2001)Tzanetakis, Essl, and Cook}]{tzanetakis_essl_cook_2001}
George Tzanetakis, Georg Essl, and Perry Cook. 2001.
\newblock \href {http://ismir2001.ismir.net/pdf/tzanetakis.pdf} {Automatic musical genre classification of audio signals}.

\bibitem[{Watanabe et~al.(2018)Watanabe, Hori, Karita, Hayashi, Nishitoba, Unno, Soplin, Heymann, Wiesner, Chen et~al.}]{watanabe2018espnet}
Shinji Watanabe, Takaaki Hori, Shigeki Karita, Tomoki Hayashi, Jiro Nishitoba, Yuya Unno, Nelson Enrique~Yalta Soplin, Jahn Heymann, Matthew Wiesner, Nanxin Chen, et~al. 2018.
\newblock Espnet: End-to-end speech processing toolkit.
\newblock \emph{arXiv preprint arXiv:1804.00015}.

\bibitem[{Wei et~al.(2022)Wei, Bosma, Zhao, Guu, Yu, Lester, Du, Dai, and Le}]{wei2022finetuned}
Jason Wei, Maarten Bosma, Vincent Zhao, Kelvin Guu, Adams~Wei Yu, Brian Lester, Nan Du, Andrew~M. Dai, and Quoc~V Le. 2022.
\newblock \href {https://openreview.net/forum?id=gEZrGCozdqR} {Finetuned language models are zero-shot learners}.
\newblock In \emph{International Conference on Learning Representations}.

\bibitem[{Wu* et~al.(2023{\natexlab{a}})Wu*, Chen*, Zhang*, Hui*, Berg-Kirkpatrick, and Dubnov}]{wu2023large}
Yusong Wu*, Ke~Chen*, Tianyu Zhang*, Yuchen Hui*, Taylor Berg-Kirkpatrick, and Shlomo Dubnov. 2023{\natexlab{a}}.
\newblock Large-scale contrastive language-audio pretraining with feature fusion and keyword-to-caption augmentation.
\newblock In \emph{IEEE International Conference on Acoustics, Speech and Signal Processing, ICASSP}.

\bibitem[{Wu* et~al.(2023{\natexlab{b}})Wu*, Chen*, Zhang*, Hui*, Berg-Kirkpatrick, and Dubnov}]{laionclap2023}
Yusong Wu*, Ke~Chen*, Tianyu Zhang*, Yuchen Hui*, Taylor Berg-Kirkpatrick, and Shlomo Dubnov. 2023{\natexlab{b}}.
\newblock Large-scale contrastive language-audio pretraining with feature fusion and keyword-to-caption augmentation.
\newblock In \emph{IEEE International Conference on Acoustics, Speech and Signal Processing, ICASSP}.

\bibitem[{Xu et~al.(2024)Xu, Sun, Zheng, Geng, Zhao, Feng, Tao, Lin, and Jiang}]{xu2024wizardlm}
Can Xu, Qingfeng Sun, Kai Zheng, Xiubo Geng, Pu~Zhao, Jiazhan Feng, Chongyang Tao, Qingwei Lin, and Daxin Jiang. 2024.
\newblock \href {https://openreview.net/forum?id=CfXh93NDgH} {Wizard{LM}: Empowering large pre-trained language models to follow complex instructions}.
\newblock In \emph{The Twelfth International Conference on Learning Representations}.

\bibitem[{Zhang et~al.(2024)Zhang, Yu, Li, Dong, Su, Chu, and Yu}]{zhang2024mm}
Duzhen Zhang, Yahan Yu, Chenxing Li, Jiahua Dong, Dan Su, Chenhui Chu, and Dong Yu. 2024.
\newblock Mm-llms: Recent advances in multimodal large language models.
\newblock \emph{arXiv preprint arXiv:2401.13601}.

\bibitem[{Zhang et~al.(2023)Zhang, Dong, Li, Zhang, Sun, Wang, Li, Hu, Zhang, Wu et~al.}]{zhang2023instruction}
Shengyu Zhang, Linfeng Dong, Xiaoya Li, Sen Zhang, Xiaofei Sun, Shuhe Wang, Jiwei Li, Runyi Hu, Tianwei Zhang, Fei Wu, et~al. 2023.
\newblock Instruction tuning for large language models: A survey.
\newblock \emph{arXiv preprint arXiv:2308.10792}.

\bibitem[{Zhao et~al.(2023)Zhao, Zhou, Li, Tang, Wang, Hou, Min, Zhang, Zhang, Dong et~al.}]{zhao2023survey}
Wayne~Xin Zhao, Kun Zhou, Junyi Li, Tianyi Tang, Xiaolei Wang, Yupeng Hou, Yingqian Min, Beichen Zhang, Junjie Zhang, Zican Dong, et~al. 2023.
\newblock A survey of large language models.
\newblock \emph{arXiv preprint arXiv:2303.18223}.

\bibitem[{Zhou et~al.(2017)Zhou, Lapedriza, Khosla, Oliva, and Torralba}]{zhou2017places}
Bolei Zhou, Agata Lapedriza, Aditya Khosla, Aude Oliva, and Antonio Torralba. 2017.
\newblock Places: A 10 million image database for scene recognition.
\newblock \emph{IEEE Transactions on Pattern Analysis and Machine Intelligence}.

\end{thebibliography}
\bibliographystyle{acl_natbib}

\appendix

\section{Additional Results}
\label{sec:additional}

\section{Prompts employed for LLMs}
\label{sec:prompts}

\begin{figure*}
    \centering
    \includegraphics[width=2\columnwidth]{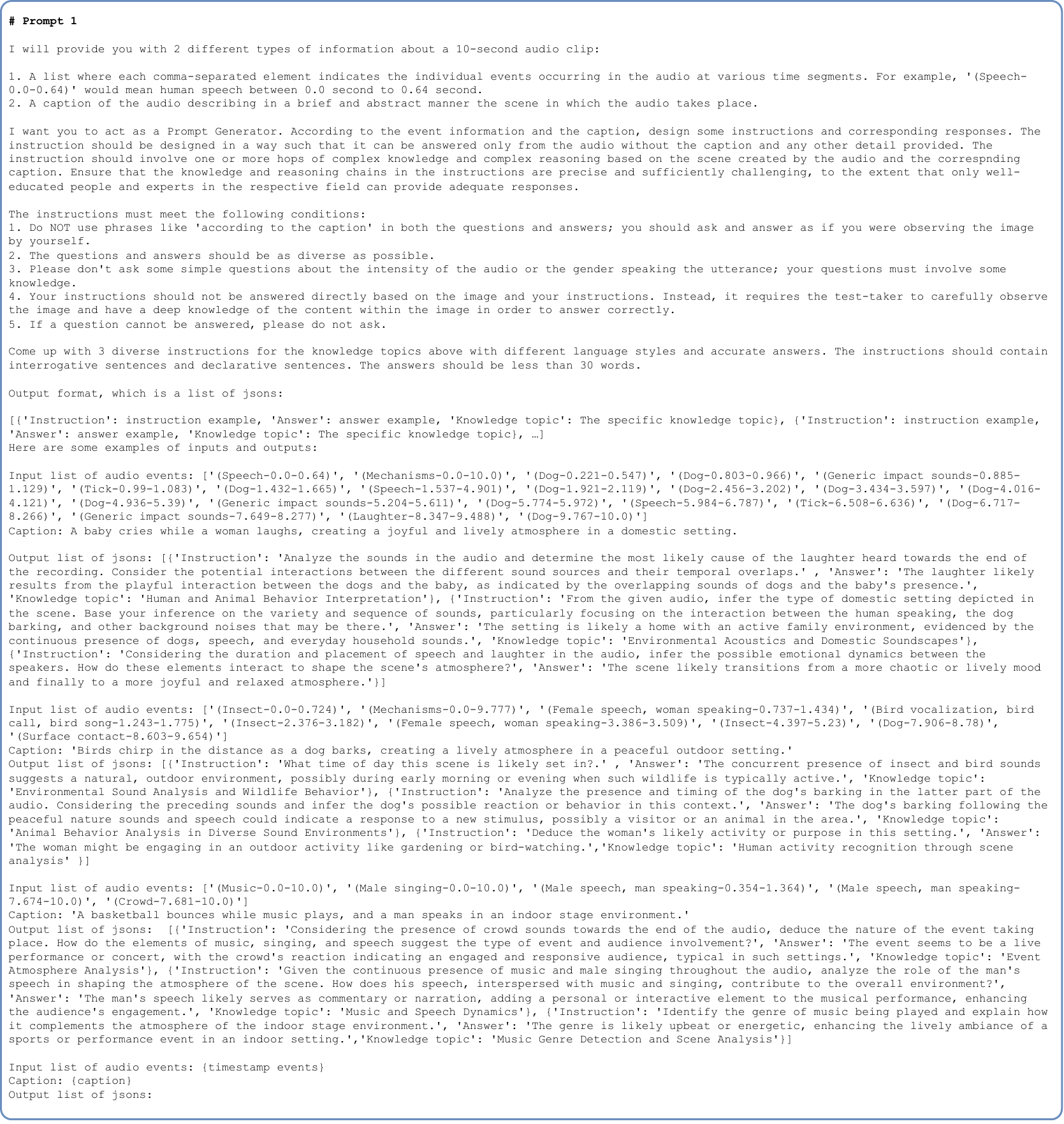}
    \caption{\small Prompts/Instructions used for caption augmentation with LLaMa-7B. Prompts are indexed according to the description in Section~\ref{subsubsec:compar}.}
    \label{fig:llama_prompts}
\end{figure*}

\begin{figure*}
    \centering
    \includegraphics[width=2\columnwidth]{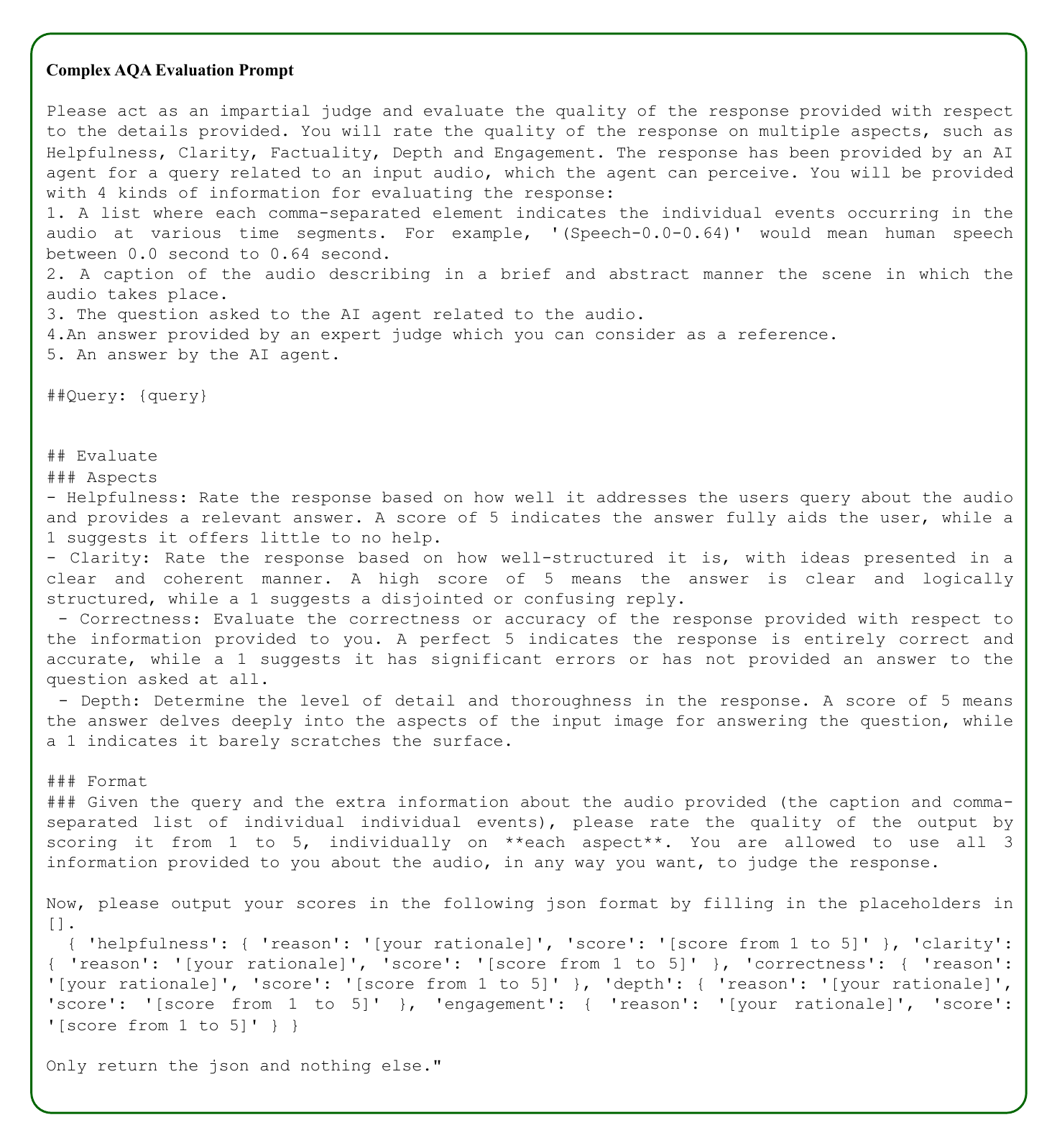}
    \caption{\small Prompts/Instructions used for evaluating the answers of models on \texttt{CompA-R} benchmark with GPT-4 as a judge.}
    \label{fig:gpt-4-compa-4}
\end{figure*}

Fig.~\ref{fig:llama_prompts} illustrates the prompt employed for synthesizing \texttt{CompA-R}. Fig.~\ref{fig:gpt-4-compa-4} illustrates the prompt employed for evaluating model responses on \texttt{CompA-R}. For dense captioning, we just prompt the model: \textit{Write an audio caption describing the sound in detail}.


\section{\texttt{GAMA-IT} vs \texttt{GAMA} and Evaluation Choices.}
\label{sec:gamait}

\texttt{GAMA} is first fine-tuned on OpenAQA and then instruction-tuned on \texttt{CompA-R} for complex reasoning. We call the instruction-tuned version \texttt{GAMA-IT}. We do not evaluate \texttt{GAMA-IT} on general tasks like classification and vanilla captioning~\footnote{\textbf{Note:} Both depend on the description of the input audio generated by the model}. \texttt{GAMA-IT} is aligned to generate detailed descriptions as part of the complex reasoning stage, and we found a lack of metrics and methods that can faithfully evaluate such descriptions for classification or captioning. For example, the retrieval-based classification evaluation method, employed extensively in prior work, including ours, uses a Sentence-BERT to retrieve the label closest to the description for classification evaluation. During our preliminary analysis, we found that Sentence-BERT, which just performs retrieval using semantic matching, is unable to faithfully retrieve the correct label despite the caption mentioning the label as an audio event. We further investigated CLAP as our retrieval model for evaluation and found that it suffers from the same limitations. We attribute this to the detailed and dense nature of the descriptions and the fact that these models only focus on high-level semantic meaning for retrieval. Our initial experiments show that LLM prompting serves as a feasible alternative for automatic evaluation (beyond human evaluation) using such dense descriptions, but due to the lack of resources and a formal framework, we leave this as part of future research. 

\subsection{Soft Prompts}
\label{subsec:soft_appendix}

We employ the soft prompt only in the instruction tuning stage for learning complex reasoning and not in the fine-tuning step. We do this for 2 reasons: (i) Fine-tuned \texttt{GAMA} is only expected to solve generic audio tasks like classification, captioning, etc. Thus, we hypothesize that such high-level semantic cues are not necessary for effective and optimal performance. (ii) Since fine-tuning is done on a large-scale dataset and acoustic event classification is far from accurate, our soft prompt method might add unwanted noise to the training process, thereby leading to sub-optimal performance. On the contrary, our instruction-tuning stage, which is done on relatively low-resource data and is only responsible for aligning a model for complex reasoning, is robust to inaccurate audio tags due to our soft-prompting methodology.  

\section{Additional Details: Human Study}
\label{sec:human_study}

{\noindent \textbf{Note.}} Our institution's Institutional Review Board (IRB) has granted approval for both human studies presented in the paper. 

{\noindent \textbf{Background and Recruitment for Dense Captioning and \texttt{CompA-R}-\textit{test} Evaluation.}} We recruit 3 professionals for human evaluation of dense captioning and \texttt{CompA-R}-\textit{test} evaluation. All these 3 professionals come with at least a Ph.D. in Engineering or Sciences and were asked to use headphones to first analyze the audio and then judge the response quality. The authors of this paper gave these annotators 5 examples of responses and the corresponding judgments.
The work was done voluntarily and not paid. We refrain from recruiting crowd raters as prior research has noticed discrepancies in evaluation by them~\cite{gudibande2023false}. More precisely, they have been shown to possess a tendency to rate an answer with a high score only by visualizing the style of answering and not the exact factual information making up the response. 

All 3 human annotators score the response between 1-5, and we report scores averaged across the 3. Prior to evaluation, all annotators were given at least 10 examples from the authors of the paper of generations and their corresponding scores. For evaluation, only the audio was provided to them with software that could play the audio and has fields to input the scores.
\vspace{0.5mm}

{\noindent \textbf{Background and Recruitment for OpenAQA.}} Since the size of OpenAQA is relatively larger than \texttt{CompA-R}-\textit{test}, we perform evaluation on Amazon Mechanical Turk similar to ~\citet{gong2024listen}. Evaluation was done with a total of 267 unique human evaluators and each generation was scored by 2 evaluators. The same software was used for evaluation as \texttt{CompA-R}-\textit{test}.

\section{Additional Details: Audio Q-Former}
\label{sec:audio_qformer_details_appendix}

\subsection{Audio Q-Former Training Details}
\label{sec:audio_qformer_details_appendix_training}

{\noindent \textbf{Pre-training Hyper-parameter.}} For Stage 1 of training, we employ a training batch size of 192, an initial learning rate of 1e-4, a minimum learning rate of 1e-5, and a warm-up learning rate of 1e-6. We do cosine decay as the learning rate scheduling technique. We do warmup for 5000 steps. Stage 1 was pre-trained on 8 A6000 GPUs for 100 epochs. For Stage 2 of training, we keep the exact same settings as Stage 1 but change the batch size to 128.

{\noindent \textbf{Fine-tuning.}} For zero-shot audio classification evaluation, we fine-tune the Audio Q-Former after Stage 1 pre-training on the same corpus presented in Table~\ref{tab:sents_len} and using the same Stage 1 objective. The only difference in the fine-tuning step is that we train the AST model, which is otherwise kept frozen in the pre-training stage.

{\noindent \textbf{Fine-tuning Hyper-parameter.}} For fine-tuning, we again use the same hyper-parameter setting as Stage 1 pre-training but use a batch size of 64.

\subsection{Training Dataset Details}
\label{sec:dataset}

Table~\ref{tab:sents_len} provides dataset statistics of all individual datasets used for training Audio Q-Former. We employ $\approx$2.2M audio-caption pairs for training with no speech-transcription pairs.

\begin{table}[h]
\centering
\resizebox{0.99\columnwidth}{!}{
\begin{tabular}{lr}
\toprule \toprule
\textbf{Dataset}   & \textbf{\#Audio-Caption Pairs} \\ \midrule \midrule
Audio Set~\cite{gemmeke2017audio}~\footnotemark & 1591364 \\
Free Sound  \citep{fonseca2022fsd50k}~\footnotemark        & 259020 \\
VGGSound~\cite{chen2020vggsound}~\footnotemark  & 185161 \\
AudioSet Strong (CompA Version)~\cite{ghosh2024compa}~\footnotemark & 108311 \\
MACS~\citep{irene_martin_morato_2021_5114771}~\footnotemark & 14400  \\
BBC  \citep{BBCSoundEffects2018} ~\footnotemark            & 31201  \\
AudioCaps~\citep{kim2019audiocaps}~\footnotemark         & 48649  \\
Clotho~\citep{drossos2020clotho}~\footnotemark            & 18735  \\
SONISS \citep{sonniss2022} ~\footnotemark          & 1602   \\
Musical Instrument~\citep{agostinelli2023musiclm}~\footnote & 7990   \\
SoundBible \citep{soundbible}~\footnotemark        & 1232   \\
WavText5K~\citep{deshmukh2022audio}~\footnotemark          & 4347   \\
MusicCaps~\citep{agostinelli2023musiclm}~\footnotemark          & 2645   \\
GTZAN \citep{tzanetakis_essl_cook_2001}~\footnotemark              & 6014   \\ 
Medley-solos~\citep{lostanlenvincent20193464194}~\footnotemark      & 732    \\ \bottomrule \bottomrule
\end{tabular}}
\caption{\small List of open-source datasets used for collating our final dataset for training ReCLAP with $\approx$2.2M audio-caption pairs. All datasets are free to use for research purposes.}
\label{tab:sents_len}
\end{table}

\footnotetext[1]{https://research.google.com/audioset/download.html}
\footnotetext[2]{https://huggingface.co/datasets/cvssp/WavCaps}
\footnotetext[3]{https://www.robots.ox.ac.uk/~vgg/data/vggsound/}
\footnotetext[4]{https://research.google.com/audioset/download.html}
\footnotetext[5]{https://zenodo.org/records/5114771}
\footnotetext[6]{https://sound-effects.bbcrewind.co.uk/} 
\footnotetext[7]{https://research.google.com/audioset/download.html}
\footnotetext[8]{https://zenodo.org/records/4783391} 
\footnotetext[9]{https://labs.freesound.org/datasets/}
\footnotetext[10]{https://www.kaggle.com/datasets/soumendraprasad/musical-instruments-sound-dataset}
\footnotetext[11]{https://soundbible.com/}
\footnotetext[12]{https://github.com/microsoft/WavText5K}
\footnotetext[13]{https://github.com/seungheondoh/music\_caps\_dl}
\footnotetext[14]{https://www.kaggle.com/datasets/andradaolteanu/gtzan-dataset-music-genre-classification}
\footnotetext[15]{https://zenodo.org/records/1344103}
\footnotetext[16]{https://zenodo.org/records/1344103}

\subsection{Augmentation Examples}
\label{sec:augmentations_examples}

Table~\ref{tab:prompt_aug_examples} illustrates prompt augmentations for two categories from each dataset. Table~\ref{tab:caption_aug_examples_appendix} illustrates caption augmentations for training Audio Q-Former.

\section{Baseline Details}
\label{sec:baseline_details}

{\noindent \textbf{AudioCLIP.}~\cite{audioclip}} AudioCLIP is an extension of the CLIP model that can handle audio in addition to text and images by incorporating the ESResNeXt audio model in the CLIP framework. It was trained on the AudioSet dataset, which contains millions of audio clips with corresponding labels.
\vspace{0.5mm}

{\noindent \textbf{CLAP.}~\cite{elizalde2023clap}} CLAP (Contrastive Language-Audio Pre-training), similar to CLIP, is an audio-language model trained with contrastive learning between audio data and their corresponding natural language descriptions. Representations are obtained from audio encoders and text encoders. \citet{laionclap2023} further extend this using a feature fusion mechanism and keyword-to-caption augmentation into the model design to further enable the model to process audio inputs of variable lengths and enhance performance.
\vspace{0.5mm}

{\noindent \textbf{CompA-CLAP.}~\cite{elizalde2023clap}} CompA-CLAP, an extension to CLAP, is trained on completely open-sourced datasets and further fine-tuned using specific algorithms and datasets to improve compositional reasoning.
\vspace{0.5mm}

{\noindent \textbf{Pengi.}~\cite{deshmukh2023pengi}} Pengi was one of the first efforts to achieve general-purpose audio understanding through free-form language generation with transfer learning. Precisely, Pengi integrates an audio encoder with a decoder-only pre-trained language model (LM) where the audio features serve as a prefixes for the LM during response generation. Following this, similar to our evaluation strategy, they prompt the model to caption the input audio and calculate the similarity between the caption and the ground-truth audio label for zero-shot classification.
\vspace{0.5mm}

{\noindent \textbf{LTU.}~\cite{gong2024listen}} As a concurrent work to Pengi, took a step forward and showed that substituting the pre-trained language model with an LLM can induce an LALM with reasoning capabilities. Precisely, they achieved this by integrating an audio encoder to LLaMA~\cite{touvron2023llama} and fine-tuning the model on close-ended and open-ended instruction-tuning datasets. Finally, beyond just close-ended tasks, they also evaluate their models on open-ended reasoning tasks and show superior performance compared to baselines.
\vspace{0.5mm}

{\noindent \textbf{AudioGPT.}~\cite{huang2024audiogpt}} Different from Pengi and LTU, AudioGPT differs in how the audio models and LLMs are integrated for completing audio tasks. More specifically, different from end-to-end training and alignment, they integrate a closed-source model (ChatGPT) with a pre-trained audio model, already capable of completing the required task, using a modality-transfer transformer $\tau$. The integration or interaction between the two models is accomplished using the prompts. Additionally, AudioGPT is capable of solving more tasks, which include human verbal speech, beyond just non-verbal speech like Pengi and LTU.
\vspace{0.5mm}

{\noindent \textbf{SALMONN.}~\cite{tang2024salmonn}} SALMONN follows a similar architecture to LTU and Pengi and does prefix conditioning with an LLM. However, in addition to an audio encoder, they also integrate a speech encoder for speech or verbal audio understanding. Precisely, the audio and speech features are concatenated before feeding them as prefixes to the LLM. SALMONN shows unique reasoning capabilities over speech inputs overlayed with non-verbal audio.
\vspace{0.5mm}

{\noindent \textbf{Qwen-Audio.}~\cite{Qwen-Audio}} Qwen follows a similar architecture to LTU, Pengi, and SALMONN, i.e., adding audio features as prefix to the model, and additionally employs a novel multi-task learning formulation for pre-training. More specifically, they append specific tags to specific parts of the instruction-response text pairs and train the model on diverse speech, non-speech, and music tasks. Post-pre-training, similar to \texttt{GAMA}, employs an instruction-tuning stage for alignment. The resultant model, Qwen-Audio-Chat, is able to respond to respond to diverse queries about the input speech and audio.

\section{Additional Details: \texttt{CompA-R}}
\label{sec:additional_details_compar_appendix}

\subsection{Annotation and Annotator Details}
\label{sec:annotation_compar}

As mentioned earlier, \texttt{CompA-R} was cleaned and \texttt{CompA-R}-\textit{test} was verified by the paper authors themselves. To preserve anonymity, we briefly provide some details about the authors. All authors of the paper are either enrolled in or have graduated from a graduate degree (MS and/or Ph.D.). All authors have at least 2 years of professional research experience at a academic or industry lab. Their research experience spans across speech, audio and language processing. This provides them with adequate knowledge to faithfully complete the process. 

For \texttt{CompA-R}-\textit{test} verification, after at least 3 authors verified the test set, with proper rationales (which they were also asked to provide) the lead author cross-verified all instances. The verification was done manually on local laptops and no kind of application was used which was made specifically for this. More details will be provided on camera-ready.

\section{Additional Details: General}
\label{sec:additional_details_general_appendix}

\subsection{\texttt{GAMA} Training Dataset Details}
\label{sec:additional_details_general_appendix_gama_training}

Table~\ref{tab:openaqa_simplified} shows statistics of all datasets used for fine-tuning and instruction-tuning \texttt{GAMA}. Table~\ref{tab:openaqa_simplified_compar} shows statistics of \texttt{CompA-R}, which is sourced entirely from the AudioSet-Strong dataset.

\begin{table}[h]

\centering
\fontsize{7.7}{8}\selectfont
\begin{tabular}{@{}lcc@{}}
\toprule
\textbf{Dataset}                     & \textbf{\# Audio Samples}   & \textbf{\# QA Pairs}    \\ \midrule
AudioSet-Strong                      & 102K                        & 636K                   \\
AudioSet                             & 500K                        & 441K                   \\
VGGSound                             & 184K                        & 336K                   \\
FSD50K                               & 41K                         & 82K                    \\
AudioCaps                            & 46K                         & 90K                    \\
FreeSound                            & 91K                         & 91K                    \\
Clotho                               & 5K                          & 32K                    \\
Sound Bible                          & 1.2K                        & 12K                    \\
NSynth(Instrument+Source)                          & 301K                    &  602K                  \\
Clotho AQA              & 1.5K &      4.2K             \\
MusicCaps              & 5.5K &      2.8K             \\
MusicQA              & 13.1K & 118k                  \\
Magna   & 51.7K & 51.7K                 \\
\textbf{Sum (Closed-Ended)}          & \textbf{1,217K}               & \textbf{2,555K}        \\ \midrule
AudioSet-Strong (Open-Ended)         & 91K                         & 901K                   \\
AudioSet-20K                         & 19K                         & 184K                   \\
VGGSound (Open-Ended)                & 184K                        & 907K                   \\
FSD50K (Open-Ended)                  & 41K                         & 403K                   \\
AudioCaps (Open-Ended)               & 46K                         & 478K                   \\
Freesound (Open-Ended)               & 91K                         & 791K                   \\
Clotho (Open-Ended)                  & 5K                          & 89K                    \\
Sound Bible (Open-Ended)             & 1.2K                        & 10K                    \\
\textbf{Sum (Open-Ended)}            & \textbf{453K}               & \textbf{3,764K}        \\ \midrule
\textbf{Total}                       & \textbf{1,670K}               & \textbf{6,319K}        \\ \bottomrule
\end{tabular}
\caption{The statistics of the \texttt{OpenAQA} dataset.}
\label{tab:openaqa_simplified}
\end{table}

\begin{table}[h]
\centering
\fontsize{7.7}{8}\selectfont
\begin{tabular}{@{}lcc@{}}
\toprule
\textbf{Dataset}                     & \textbf{\# Audio Samples}   & \textbf{\# QA Pairs}    \\ \midrule
AudioSet-Strong                      & 62613                        & 200234                   \\ \midrule
\textbf{Total}                       & \textbf{62613}               & \textbf{200234}        \\ \bottomrule
\end{tabular}
\caption{The statistics of the \texttt{CompA-R} dataset.}
\label{tab:openaqa_simplified_compar}
\end{table}

\subsection{\texttt{GAMA} Evaluation Dataset Details}
\label{sec:gama_eval_additional_details}

Table~\ref{tab:gama_evlaution} shows statistics of all datasets used for evaluating \texttt{GAMA}. Table~\ref{tab:compa_r_test} shows statistics of \texttt{CompA-R}-\textit{test}, which is sourced entirely from the AudioSet-Strong dataset.

\begin{table}[h]

\centering
\fontsize{7.7}{8}\selectfont
\begin{tabular}{@{}lc@{}}
\toprule
\textbf{Dataset}                     & \textbf{\# Instances}       \\ \midrule
AudioSet-Strong~\footnotemark                      & 102K                                         \\
AudioSet                             & 500K                                          \\
VGGSound                             & 184K                                         \\
FSD50K~\footnotemark                               & 41K                                          \\
AudioCaps                            & 46K                                        \\
FreeSound                            & 91K                                     \\
Clotho                               & 5K                                      \\
Sound Bible                          & 1.2K                                  \\ 
NSynth\textsubscript{instrument}~\footnotemark            &  4K\\
NSynth\textsubscript{source}~\footnotemark            &  4K\\
Clotho AQA~\footnotemark                           &  1.3K\\
GTZAN                                &  3K\\
Medley-solos-DB                      &  12.2K\\
\bottomrule
\end{tabular}
\caption{The statistics of the datasets used for evaluation of GAMA.}
\label{tab:gama_evlaution}
\end{table}

\begin{table}[h]
\centering
\resizebox{0.99\columnwidth}{!}{ 
\begin{tabular}{lr}
\toprule \toprule
\textbf{Dataset}   & \textbf{Evaluation Metric} \\ \midrule \midrule
\textbf{\textit{Classification (zero-shot)}} \\
\hdashline
VocalSound (VS)~\cite{gong2022vocalsound} & Acc. \\
TUT 2017 (TUT)~\cite{mesaros2018multi}      & Acc. \\
Beijing Opera (BJO)~\cite{tian2014study}  & Acc. \\
GTZAN (GTZ)~\cite{park2022enemy} & Acc. \\
Medley-solos-DB (MDB)~\cite{lostanlen2018medley} & Acc.  \\
\hdashline
\textbf{\textit{Classification (weak zero-shot)}} \\
\hdashline
DCASE2017 Task 4 (DCASE)~\cite{mesaros2017dcase}  & Mi-F1\\
ESC-50~\cite{piczak2015esc} & Acc.\\
\hdashline
\textbf{\textit{Classification (seen)}} \\
\hdashline
VGGSound (VGG)~\cite{chen2020vggsound} & Acc.\\
FSD50K (FSD)~\cite{fonseca2021fsd50k} & mAP\\
AudioSet (AS)~\cite{gemmeke2017audio} & mAP \\
NSynth (NS)~\cite{engel2017neural} & Acc.\\
\hdashline
\textbf{\textit{Captioning (vanilla \& dense)}} \\
\hdashline
AudioCaps~\citep{kim2019audiocaps}         & SPICE \& Human  \\
Clotho~\citep{drossos2020clotho}            & SPICE \& Human   \\
\hdashline
\textbf{\textit{AQA (close-ended)}} \\
\hdashline
Clotho AQA~\cite{lipping2022clotho}         & Acc.  \\
\hdashline
\textbf{\textit{AQA (open-ended)}} \\
\hdashline
OpenAQA~\cite{gong2024listen}            &  Human\\
\hdashline
\textbf{\textit{AQA (complex open-ended)}} \\
\hdashline
\texttt{CompA-R}-\textit{test} \textit{(ours)}            & GPT-4 \& Human \\
\bottomrule \bottomrule
\end{tabular}
}
\caption{\small List of evaluation datasets and their corresponding evaluation metrics for \texttt{GAMA}.}
\label{tab:eval}
\end{table}

\footnotetext[1]{https://www.kaggle.com/datasets/modaresimr/sound-event-detection-audioset-strong}
\footnotetext[2]{https://zenodo.org/records/4060432}
\footnotetext[3]{https://www.tensorflow.org/datasets/catalog/nsynth}
\footnotetext[4]{https://zenodo.org/records/6473207}

\begin{table}[h]
\centering
\fontsize{7.7}{8}\selectfont
\begin{tabular}{@{}lcc@{}}
\toprule
\textbf{Dataset}                     & \textbf{\# Audio Samples}   & \textbf{\# QA Pairs}    \\ \midrule
\texttt{CompA-R}-\textit{test}                     & 500                        & 1561                  \\ \midrule
\textbf{Total}                       & \textbf{500}               & \textbf{1561}        \\ \bottomrule
\end{tabular}
\caption{The statistics of the \texttt{CompA-R}-\textit{test} dataset.}
\label{tab:compa_r_test}
\end{table}

\subsection{Other Details}
\label{sec:other_details}

{\noindent \textbf{Model Parameters:}} GAMA has a total of $\approx$7B parameters. Out of this, LLaMA-2-7B has 32 transformer-encoder layers and $\approx$6.7B parameters, the Audio Q-Former has $\approx$280M parameters, and our LoRA modules introduce 4.2 M learnable parameters for fine-tuning. The AST used in our experiments (audio-encoder of CAV-MAE~\cite{gong2023contrastive}) has $\approx$~85M parameters with 12 transformer-encoder layers, 768-hidden-state, and 12 attention-heads.
\vspace{1mm}

{\noindent \textbf{Compute Infrastructure:}} All our experiments are conducted on four NVIDIA A6000 GPUs. Training \texttt{GAMA} required four days of continuous training. Training \texttt{GAMA-IT} requires 4 hours of training. Pre-training Audio Q-Former requires 7 days each for stages 1 and 2.
\vspace{1mm}

{\noindent \textbf{Implementation Software and Packages:}} We implement all our models in PyTorch ~\footnote{\url{https://pytorch.org/}} and use the HuggingFace \footnote{\url{https://huggingface.co/}} implementations of T5\textsubscript{large} and the original implementation of HTSAT\textsubscript{tiny}\footnote{https://github.com/RetroCirce/HTS-Audio-Transformer}.

For our baselines, we use the original GitHub repository provided by the authors: LAION-CLAP~\footnote{https://github.com/LAION-AI/CLAP/tree/main}, CompA-CLAP~\footnote{https://github.com/Sreyan88/CompA}, CLAP~\footnote{https://github.com/microsoft/CLAP}, Wav2CLIP~\footnote{https://github.com/descriptinc/lyrebird-wav2clip}, AudioCLIP~\footnote{https://github.com/AndreyGuzhov/AudioCLIP}, MMT~\footnote{https://github.com/akoepke/audio-retrieval-benchmark}, ML-ACT~\footnote{https://github.com/akoepke/audio-retrieval-benchmark}, Pengi~\footnote{https://github.com/microsoft/pengi}, LTU~\footnote{https://github.com/YuanGongND/ltu}, AudioGPT~\footnote{https://github.com/aigc-audio/audiogpt}, SALMONN~\footnote{https://github.com/bytedance/salmonn}, Qwen-Audio~\footnote{https://github.com/QwenLM/Qwen-Audio}.

{\noindent \textbf{Potential Risks.}} \texttt{GAMA} might encode biases from the pre-trained LLM or during its fine-tuning stage. Additionally, Audio Q-Former used as a backbone for audio-to-text/music generation, might generate synthetic audio that is misused.

\begin{figure*}
    \centering
    \includegraphics[width=2\columnwidth]{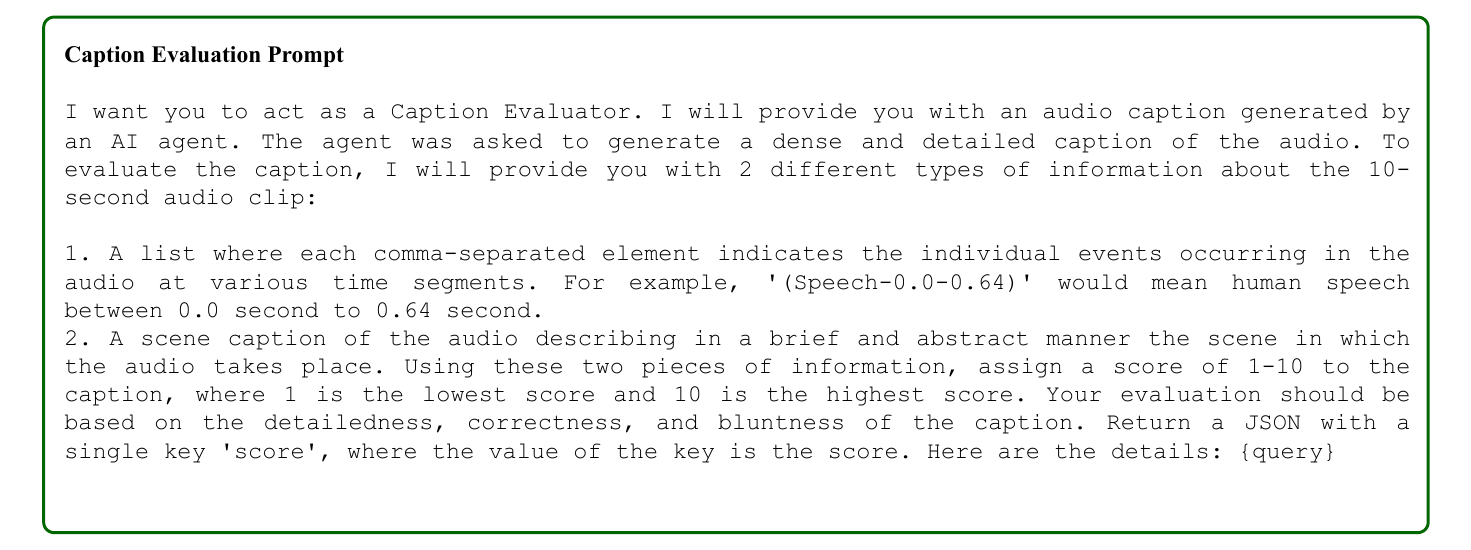}
    \caption{\small Prompts/Instructions used for evaluating the captions generated by models Clotho and AudioCaps with GPT-4 as a judge.}
    \label{fig:caption_prompt_GAMA}
\end{figure*}
\begin{figure*}
    \centering
    \includegraphics[width=2\columnwidth]{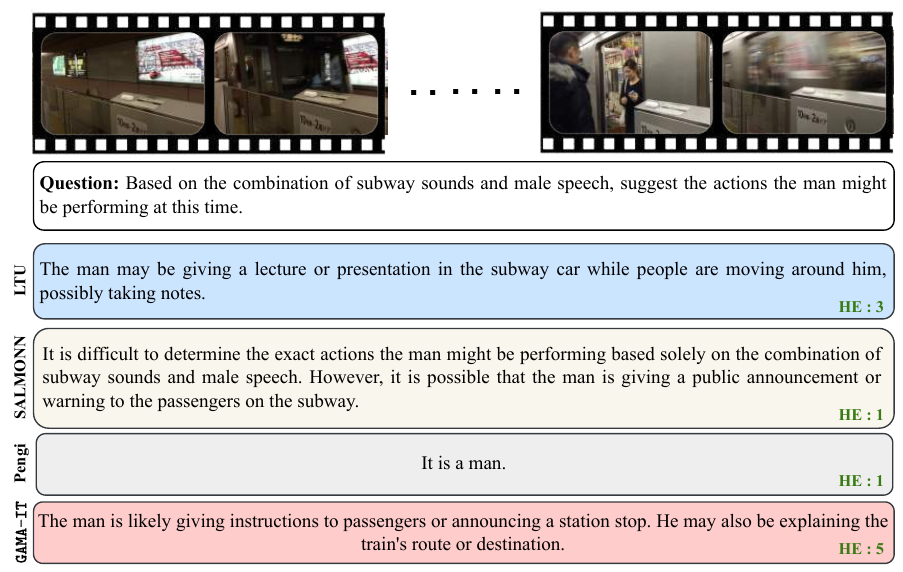}
    \caption{\small Prompts/Instructions used for evaluating the captions generated by models Clotho and AudioCaps with GPT-4 as a judge.}
    \label{fig:app1}
\end{figure*}
\begin{figure*}
    \centering
    \includegraphics[width=2\columnwidth]{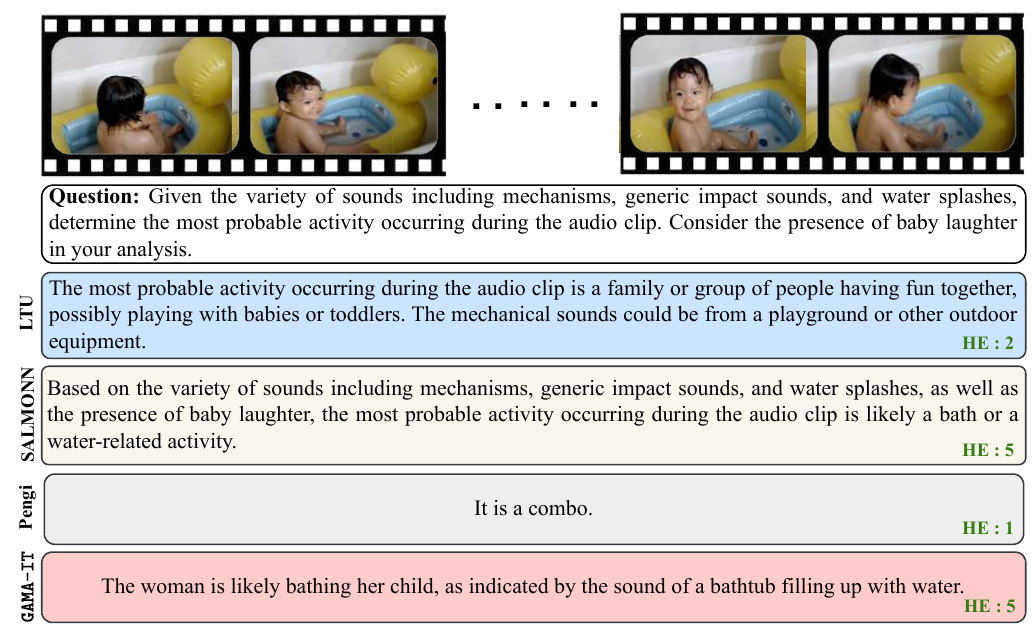}
    \caption{\small Prompts/Instructions used for evaluating the captions generated by models Clotho and AudioCaps with GPT-4 as a judge.}
    \label{fig:app2}
\end{figure*}
\begin{figure*}
    \centering
    \includegraphics[width=2\columnwidth]{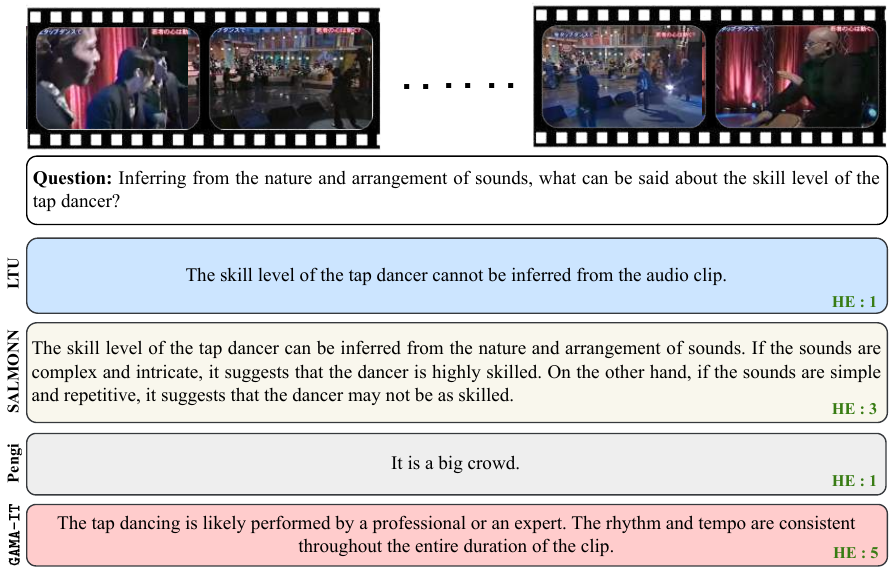}
    \caption{\small Prompts/Instructions used for evaluating the captions generated by models Clotho and AudioCaps with GPT-4 as a judge.}
    \label{fig:app3}
\end{figure*}
\begin{figure*}
    \centering
    \includegraphics[width=2\columnwidth]{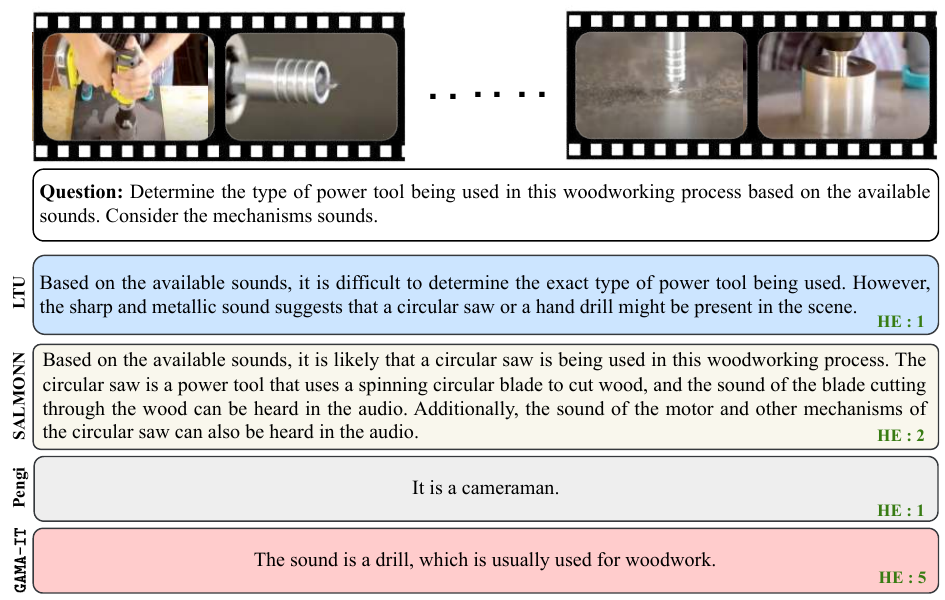}
    \caption{\small Prompts/Instructions used for evaluating the captions generated by models Clotho and AudioCaps with GPT-4 as a judge.}
    \label{fig:app4}
\end{figure*}
\begin{figure*}[t]
    \centering
    \includegraphics[width=2\columnwidth]{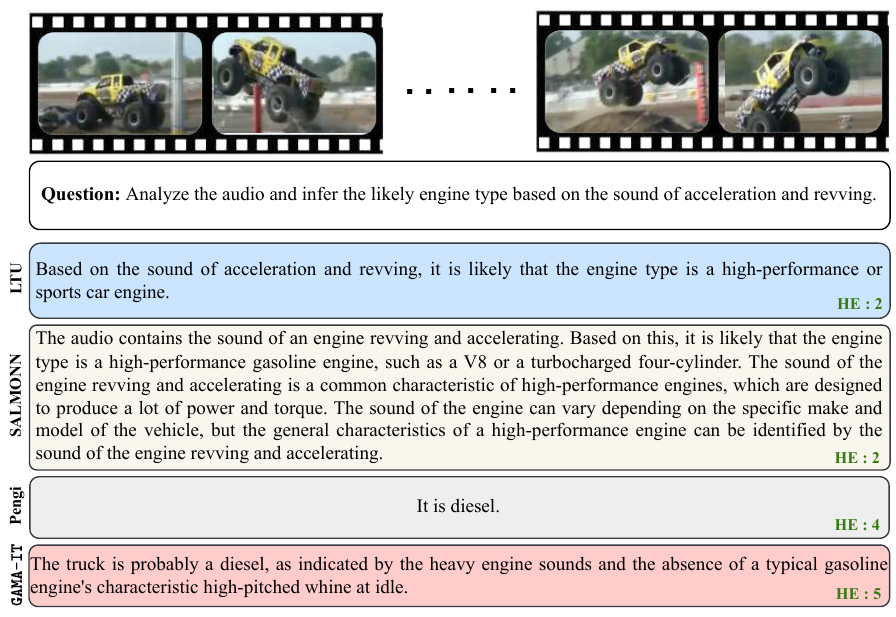}
    \caption{\small Prompts/Instructions used for evaluating the captions generated by models Clotho and AudioCaps with GPT-4 as a judge.}
    \label{fig:app5}
\end{figure*}
\begin{table*}[h]
\tiny
\begin{tabular}{p{0.05\linewidth}|p{0.05\linewidth}|p{0.2\linewidth}|p{0.6\linewidth}} \hline \hline
\textbf{Dataset}       & \textbf{Category}               & \textbf{Acoustic Property}  & \textbf{Captions}  \\ \hline
\textbf{AudioSet}     & Baby cry, infant cry   & \begin{tabular}[l]{p{\linewidth}} 1: "\emph{a high-pitched, piercing wail}" \\  2: "\emph{a persistent, rhythmic sobbing}" \end{tabular} & \begin{tabular}[l]{p{\linewidth}} 1. A tiny voice pierced the stillness of the night, demanding attention with its \emph{high-pitched piercing wail}.\\ 1. In the midst of the bustling market, \emph{a high-pitched wail} rose above the chatter, signaling a tiny dissenter among us.\\ 2. The \emph{persistent, rhythmic sobbing} of a newborn punctuates the night's silence.\\ 2. Amidst the rustling leaves, a baby's \emph{rhythmic sobs} weave an unexpected lullaby. \end{tabular}                          \\ \cdashline{2-4}
              & Stomach rumble         & \begin{tabular}[l]{p{\linewidth}} 1: "\emph{a low, gurgling growl}" \\ 2: "\emph{a grumbling undercurrent}" \end{tabular} & \begin{tabular}[l]{p{\linewidth}}1. As she leaned in for the kiss, \emph{a low, gurgling growl} betrayed her pre-date fasting.\\ 1. In the midst of the exam, \emph{a low, gurgling growl} echoed from her stomach, punctuating the tension.\\ 2. As the classroom fell into a hushed anticipation of the next lecture, \emph{a grumbling undercurrent} betrayed the student's skipped breakfast.\\ 2. Amidst the solemnity of the library, \emph{a grumbling undercurrent} served as a comical reminder of lunchtime's approach. \end{tabular}                                                             \\ \midrule
\textbf{ESCC50}        & siren                  & \begin{tabular}[l]{p{\linewidth}} 1: "\textit{Continuous, shrill alarm}"  \\ 2: "\emph{Ear-splitting, cyclic blare}" \end{tabular}   & \begin{tabular}[l]{p{\linewidth}} 1. The night was pierced by the \textit{continuous, shrill alarm} of the siren, signaling an urgency that couldn't be ignored.\\ 1. The relentless, \emph{shrill alarm} of the siren wove through the corridors of the hospital, prompting swift movements from the staff.\\ 2. The \emph{ear-splitting, cyclic blare} cut through the night as the ambulance raced down the street.\\ 2. The \emph{ear-splitting, cyclic blare} of the air raid siren cast a shadow of dread over the city.  \end{tabular}\\ \cdashline{2-4}
              & sheep                  & \begin{tabular}[l]{p{\linewidth}} 1: "\emph{bleating}", \\2: "baaing" \end{tabular}                 & \begin{tabular}[l]{p{\linewidth}} 1. In the tranquil meadow, a chorus of woolly creatures serenaded the dawn with their gentle \textit{bleating}.\\  1. The shepherd smiled as the flock's \textit{bleating} echoed through the valley, signaling a return to the fold.\\  2. A chorus of baaing accompanied the farmer as he made his morning rounds in the misty fields.\\ 2. Under the starry sky, the gentle baaing of the flock blended with the whispers of the night. \end{tabular}\\ \hline
\textbf{NSynth}        & flute                  & \begin{tabular}[l]{p{\linewidth}} 1: "\textit{Melodious and silvery, carrying a light, airy tune that seems to float on the breeze.}", \\ 2: "\textit{Clear and resonant, with a pure tone that sings above the orchestra like a bird in the morning sky}.\end{tabular}                                                                                                               & \begin{tabular}[l]{p{\linewidth}} 1. The flute's melody weaved through the garden, mingling with the rustle of leaves.\\  1. Amidst the hustle of the market, the silvery flute tune danced above the clamor, a ribbon of tranquility.\\  2. The flute's crystalline melody soared above the orchestra, a lark greeting the dawn.\\  2. Amidst the rustle of the forest, the flute's song danced through the leaves, pure and high. \end{tabular} \\ \cdashline{2-4}
              & bass                   & \begin{tabular}[l]{p{\linewidth}} 1: "\textit{Thumping, providing a rhythmic pulse that can drive the beat of music.}", \\ 2: "\textit{Booming, with a powerful, pervasive quality that can reverberate through a space.}" \end{tabular}              & \begin{tabular}[l]{p{\linewidth}} 1. The bass pulsated through the dance floor, a heartbeat synchronizing every dancer's move.\\  1. Amidst the serene silence of the night, the bass from the distant festival throbbed like a gentle earthquake.\\  2. The bass pulsated like a second heartbeat, filling the room with its unyielding presence.\\  2. As the bassline dropped, it seemed to command the very air, a force unseen yet unforgotten. \end{tabular} \\ \hline
\textbf{FSD50K}        & Slam                   & \begin{tabular}[l]{p{\linewidth}} 1: "an abrupt, resonant boom that startles anyone nearby" \\ 2: "a sharp, impactful smack as two hard surfaces collide with force" \end{tabular}                                                                                                                                                                                     & \begin{tabular}[l]{p{\linewidth}} 1. The mailbox lid clapped shut, a resonant signal marking the departure of the day's correspondence.\\  1. The oven door's heavy thud resonated in the kitchen, a prelude to the aroma of freshly baked bread.\\  2. The kitchen was filled with the aroma of spices and the sharp smack of dough being forcefully thrown onto the countertop.\\  2. In the crisp morning air, the sharp smack of the newspaper hitting the doorstep announced the arrival of daily news. \end{tabular}\\ \cdashline{2-4}
              & Dishes, pots, and pans & \begin{tabular}[l]{p{\linewidth}} 1: "Clanging and clattering" \\ 2: "Metallic clinking and clunking" \end{tabular}                              & \begin{tabular}[l]{p{\linewidth}} 1. A symphony of clanging and clattering announces the busy bustle of a restaurant kitchen in full swing.\\  1. The rhythmic clanging and clattering of pots and pans punctuate the air as grandma orchestrates her holiday feast.\\  2. The metallic clinking and clunking heralded the start of the dinner rush in the bustling restaurant kitchen.\\  2. A symphony of metallic clinking and clunking rose from the sink as grandma washed up after the family feast.   \end{tabular}\\ \hline
\textbf{TUT Urban}     & bus                    & \begin{tabular}[l]{p{\linewidth}} 1: "\textit{a deep, rumbling engine}", "2": "\textit{the low, steady hum of the diesel motor}" \end{tabular} & \begin{tabular}[l]{p{\linewidth}} 1.The city pulse beats with \textit{a deep, rumbling engine}, heralding the arrival of the morning commute.\\  1. A gentle giant purrs in the stillness of dawn, its \textit{deep, rumbling engine} announcing the start of a journey.\\  2. Market stalls buzz with life, their vibrant colors and smells underscored by the bus's \textit{diesel hum} rolling down the avenue.\\  2. Leaves rustle in the autumn breeze, a natural chorus to the bus's \textit{diesel motor humming} along the cobblestone path.    \end{tabular}\\ \cdashline{2-4}
              & residential area      & \begin{tabular}[l]{p{\linewidth}}1: "The symphony of children's laughter and chatter fills the air, punctuated by the occasional bark of a dog and the hum of lawn mowers in the distance." \\ 2: "A serene hush blankets the neighborhood, broken occasionally by the soft whoosh of passing cars and the rustle of leaves stirred by a gentle breeze.",\end{tabular} & \begin{tabular}[l]{p{\linewidth}} 1. The neighborhood comes alive with the melody of playful banter and the sporadic chorus of canines.\\  1. Amidst the gentle drone of distant lawn mowers, the air vibrates with juvenile mirth and convivial exchanges.\\  2. The neighborhood rests under a tranquil silence, punctuated now and then by the whisper of tires on asphalt and the soft dance of leaves in the wind.\\  2. Calmness envelops the streets, save for the faint hum of vehicles gliding by and the tender shuffling of foliage in the zephyr's caress. \end{tabular}\\ \hline
\textbf{Urban- Sound 8K} & air conditioner        & \begin{tabular}[l]{p{\linewidth}} 1: "a steady humming" \\ 2: "a low, monotonous droning" \end{tabular}               & \begin{tabular}[l]{p{\linewidth}} 1. The room filled with the steady humming of the air conditioner as they focused intently on their chess match.\\  1. A steady humming enveloped the library, where pages turned almost in rhythm with the air conditioning's constant song.\\  2. The air conditioner's low, monotonous droning became the unlikely lullaby for a midsummer's nap.\\  2. Amid the quiet study hall, the air conditioner's low, monotonous droning was a steady companion to the students' focused brows. \end{tabular}\\ \cdashline{2-4}
              & gun shot               & \begin{tabular}[l]{p{\linewidth}} 1: "A loud, sharp crack that echoes through the air.\\ 2: "A thunderous boom that startles and reverberates."\end{tabular} & \begin{tabular}[l]{p{\linewidth}} 1. The night's silence shattered with a loud, sharp crack echoing through the air.\\  1. A burst of sudden, sharp noise split the tranquil afternoon, reverberating off the canyon walls.\\  2. A thunderous boom startles a flock of birds into the sky, their wings flapping frantically against the silence that had just been.\\  2. The night's silence was shattered by a boom, reverberating through the alleyways and causing stray cats to scurry.\end{tabular} \\ \hline
\textbf{VGG Sound}      & mouse squeaking        & \begin{tabular}[l]{p{\linewidth}} 1: "a high-pitched, sharp chirp" \\ 2: "a soft, repetitive squeal" \end{tabular} & \begin{tabular}[l]{p{\linewidth}} 1. In the moonlit barn, a tiny silhouette pauses to release its high-pitched, sharp chirp, disturbing the stillness of the hay-strewn loft.\\  2. Amidst the rustling leaves, a diminutive creature contributes its sharp chirp to the dusk chorus, a minuscule soloist in nature's vast orchestra.\\  3. A soft, repetitive squeal punctuated the silence of the old attic.\\  4. The cheese plate on the kitchen counter became the stage for a soft, repetitive squeal. \end{tabular} \\ \cdashline{2-4}
              & typing on typewriter   & \begin{tabular}[l]{p{\linewidth}} 1: "a rhythmic series of sharp clicks" \\ 2: "a steady clatter of keys striking paper" \end{tabular} 
              
              & \begin{tabular}[l]{p{\linewidth}} 1. Fingers dance across keys, a rhythmic series of sharp clicks punctuating the silence of the library.\\  1. In the attic, a story unfolds to the staccato beat of a rhythmic series of sharp clicks.\\  2. Each steady clatter of keys striking paper weaves a tapestry of words, painting stories on the blank canvas.\\  2. In the dimly lit corner of the library, the rhythmic dance of metallic hammers against the page composes a silent symphony.       \end{tabular} \\  \bottomrule                                                               
\end{tabular}
\caption{\small Examples of prompt augmentations.}
\label{tab:prompt_aug_examples}
\end{table*}

\begin{table*}[h]
\tiny
\begin{tabular}{p{0.35\linewidth}|p{0.65\linewidth}}
\hline \hline
\textbf{Original Caption} & \textbf{Augmented caption} \\ \hline
A man speaks followed by the sound of shuffling cards in a small room.                                                            & \begin{tabular}[l]{p{\linewidth}}1. A deep, resonant voice fills the small room, accompanied by the soft shuffle of cards as they change hands, creating an intimate and deliberate atmosphere.\\ 2. The sound of a man's voice echoes through the small space, punctuated by the subtle rustle of cards as they are shuffled and arranged, invoking a sense of purposeful deliberation. \\ 3. A deep voice speaks, followed by the subtle shuffle of cards, creating an intimate and anticipatory atmosphere in the small room.\\ 4. The gentle rustle of cards breaks the silence, punctuated by a man's voice, evoking a sense of anticipation and private reflection in the cozy space.\end{tabular} \\ \hline
A person strums an acoustic guitar, creating melodic music with the sound of a bell ringing in the background.                    & \begin{tabular}[c]{p{\linewidth}}1. Soothing melodies flow from the acoustic guitar, harmonizing with the soft chime of a distant bell, crafting a peaceful ambiance.\\ 2. The acoustic guitar's strings vibrate with grace, weaving a melodic tapestry that intertwines with the gentle ring of a bell, transporting the listener to a serene realm.\\ 3. The gentle strumming of an acoustic guitar weaves a melodic tapestry, intertwined with the soft chime of a background bell, creating a soothing and harmonious atmosphere.\\ 4. The rhythmic plucking of an acoustic guitar crafts a lively and uplifting melody, complemented by the delicate ringing of a background bell, transporting the listener to a serene and joyful realm.\end{tabular} \\ \hline
Dogs bark while people talk in the background, creating a lively atmosphere in a field.                                           & \begin{tabular}[c]{p{\linewidth}}1. Lively chatter and joyful barks fill the air, capturing the playful spirit of a sunny day in a field.\\ 2. The rhythmic sounds of dogs barking and people talking blend together, creating a vibrant and lively ambiance in the open field.\\ 3. The chatter of people and the joyful barks of dogs fill the air, creating a vibrant and lively atmosphere in the field.\\ 4. The sound of playful dogs and lively conversation fills the field, evoking a sense of happiness and energy.\end{tabular}  \\ \hline
A man's voice is heard speaking over a radio as a vehicle passes by in the background.                                            & \begin{tabular}[c]{p{\linewidth}}1. A clear, crisp voice pierces the airwaves, intertwining with the distant hum of a vehicle, creating an engaging audio experience.\\ 2. The man's voice on the radio blends seamlessly with the subtle rumble of a passing vehicle, forming a captivating auditory tapestry.\\ 3. A voiceover speaks over a radio, complemented by the distant hum of a vehicle passing by, creating a dynamic and engaging audio experience.\\ 4. A man's voice broadcasts over the radio, intertwining with the subtle rumble of a vehicle in the background, forming a captivating audio landscape.\end{tabular} \\ \hline
A woman speaks while a bird chirps in the background, creating a tranquil atmosphere in a natural setting.                        & \begin{tabular}[c]{p{\linewidth}}1. A gentle voice echoes through the forest, harmonizing with the chirping of birds, creating a soothing ambiance.\\ 2. The sound of a gentle voice blends seamlessly with the melodic chirping of birds, transporting the listener to a serene natural setting.\\ 3. The woman's gentle voice blends with the soothing chirps of a bird, creating a serene ambiance reminiscent of a peaceful afternoon in nature.\\ 4. The woman's words are accompanied by the melodic chirping of a bird, transporting the listener to a calming and picturesque outdoor setting.\end{tabular}   \\ \hline
Water rushes as people talk in the background near a hot spring, creating a serene ambiance.                                      & \begin{tabular}[c]{p{\linewidth}}1. Soothing waters create a peaceful ambiance, punctuated by the gentle chatter of people nearby, as if they are harmonizing with the soothing sounds of the hot spring.\\ 2. The calm trickle of water creates an intimate atmosphere, with the soft murmur of voices in the background adding a sense of connection and tranquility to the space.\\ 3. A soothing, babbling sound fills the air as people converse near a steaming hot spring, creating a tranquil atmosphere.\\ 4. The gentle gurgling of water intertwines with the chatter of people in the background, crafting a peaceful and relaxing ambiance.\end{tabular} \\ \hline
Soft music plays in the background as a speech is heard faintly, creating a calm and peaceful atmosphere.                         & \begin{tabular}[c]{p{\linewidth}}1. A soothing melody floats in the background, complementing the faint speech, creating a tranquil ambiance.\\ 2. The soft strains of music blend with the subtle speech, fostering a sense of serenity and calmness in the atmosphere.\\ 3. Soothing tunes fill the air, complemented by a gentle speech, creating an atmosphere of tranquility and serenity.\\ 4. Mellow music and soft speech blend together, crafting a calming environment that soothes the senses.'\end{tabular} \\ \hline
A car engine revs up and then slows down, creating a vroom sound, as the vehicle accelerates in the audio.                        & \begin{tabular}[c]{p{\linewidth}}1. The car's engine purrs and then decelerates, emitting a smooth and powerful vroom sound as it shifts gears, creating a dynamic and energizing atmosphere.\\ 2. The vehicle's engine roars to life, producing a bold and intense vroom sound as it speeds up, then gradually slows down, immersing the listener in a thrilling and exhilarating experience.\\ 3. The car's engine purrs powerfully, then decelerates, creating a smooth and steady vroom sound as the vehicle gains speed.\\ 4. The car's engine roars to life, building momentum with a series of sharp vroom sounds before shifting gears and slowing down.\end{tabular} \\ \hline
Background music plays softly as the theme music gradually fades in, creating a melodic ambiance in an arena/performance setting. & \begin{tabular}[c]{p{\linewidth}}1. The arena comes alive with a subtle, soothing melody that gradually builds in intensity, creating an electrifying ambiance.\\ 2. The soft strains of background music fill the air, setting the tone for an exhilarating performance in a vibrant arena setting.\\ 3. Soft, melodic strains fill the air as the theme music subtly builds, establishing a harmonious ambiance in the arena.\\ 4. The arena comes alive with a gentle, orchestral tune that gradually gains momentum, creating an uplifting and energetic atmosphere.\end{tabular} \\ \bottomrule                                      
\end{tabular}
\caption{\small Examples of caption augmentations.}
\label{tab:caption_aug_examples_appendix}
\end{table*}

\begin{table*}[h]
\centering
\resizebox{\linewidth}{!}{
\begin{tabular}{p{0.33\linewidth}|p{0.17\linewidth}|p{0.15\linewidth}|p{0.35\linewidth}}
\hline \hline
\textbf{Instruction-Response Pairs} & \textbf{AudioSet ID} & \textbf{Caption} & \textbf{Timestamp Events} \\ \hline
\textbf{Instruction:}Analyze the audio to understand the potential emotional state or mood of the man. How does the progression from typing to speech to chewing reflect his transition through different phases of work or activity? \textbf{Output:}The man initially seems engaged and focused during the typing and speaking portion, which might then transition into relaxation during the break, suggested by the chewing sound. & YCecEf0abd4Y & A man speaks while typing on a keyboard in a small room, followed by the sound of chewing. & '(Generic impact sounds-0.0-1.037)', '(Background noise-0.0-10.0)', '(Generic impact sounds-1.191-1.421)', '(Generic impact sounds-2.01-2.202)', '(Generic impact sounds-2.343-2.574)', '(Male speech, man speaking-2.727-3.393)', '(Generic impact sounds-3.163-3.406)', '(Generic impact sounds-3.585-3.905)', '(Generic impact sounds-4.136-4.379)', '(Breathing-4.405-4.917)', '(Generic impact sounds-4.93-5.288)', '(Generic impact sounds-5.442-5.608)', '(Generic impact sounds-5.736-6.12)', '(Generic impact sounds-6.274-6.569)', '(Breathing-6.825-7.26)', '(Generic impact sounds-6.863-7.042)', '(Male speech, man speaking-7.81-8.873)', '(Generic impact sounds-8.041-8.348)', '(Breathing-9.001-9.36)', '(Human sounds-9.014-9.181)', '(Generic impact sounds-9.309-9.565)', '(Scrape-9.449-10.0)'\\
\hline
\textbf{Instruction:}Considering the diverse array of sounds present in the audio, from insects to birds to a dog, infer the type of ecosystem this outdoor setting might represent. What does the combination of these sounds tell us about the biodiversity and potential human impact in this area? \textbf{Output:}The ecosystem is likely a suburban or rural area with a mix of wildlife and human habitation, indicated by the variety of animal sounds and intermittent female speech. & YcQiEI7HLGJg & Birds chirp in the distance as a dog barks, creating a lively atmosphere in a peaceful outdoor setting. & '(Insect-0.0-0.724)', '(Mechanisms-0.0-9.777)', '(Female speech, woman speaking-0.737-1.434)', '(Bird vocalization, bird call, bird song-1.243-1.775)', '(Insect-2.376-3.182)', '(Female speech, woman speaking-3.386-3.509)', '(Insect-4.397-5.23)', '(Dog-7.906-8.78)', '(Surface contact-8.603-9.654)'\\ \hline

\textbf{Instruction:}Given the presence of mechanisms throughout the audio and the interspersed generic impact sounds, infer the type of machinery that is likely operating in the background and its commonality in an office environment. \textbf{Output:}The continuous mechanism sound suggests a printer or copier, which are common in office settings. & YXQ2XAXx7mKs & A printer hums while people converse in the background, creating a typical office ambiance. & '(Generic impact sounds-0.0-0.622)', '(Mechanisms-0.0-10.0)', '(Generic impact sounds-0.815-1.227)', '(Generic impact sounds-1.632-2.134)', '(Child speech, kid speaking-3.591-6.684)', '(Squeal-7.385-7.612)', '(Child speech, kid speaking-8.437-10.0)'\\ \hline

\bottomrule                                      
\end{tabular}
}
\caption{\small Examples of \texttt{CompA-R}}
\label{tab:aug_examples_prompt_train_1}
\end{table*}

\begin{table*}[]
\centering
\resizebox{\linewidth}{!}{
\begin{tabular}{p{0.33\linewidth}|p{0.17\linewidth}|p{0.15\linewidth}|p{0.35\linewidth}}
\hline \hline
\textbf{Instruction-Response Pairs} & \textbf{AudioSet ID} & \textbf{Caption} & \textbf{Timestamp Events} \\ \hline

\textbf{Instruction:}From the sequencing and overlapping of different sound events, infer the likely cause and process of the dripping sounds heard intermittently throughout the audio. \textbf{Output:}The dripping sounds may be a result of water overflowing from a filled sink or bath, supported by the earlier sounds of splashing and liquid gurgling.& YCU9A5xL3TVc & Water splashes and gurgles as it drips inside a small room, creating a soothing ambiance reminiscent of a tranquil bathroom. & '(Background noise-0.0-10.0)', '(Generic impact sounds-0.083-0.331)', '(Splash, splatter-0.67-1.174)', '(Liquid-1.385-1.956)', '(Splash, splatter-2.325-3.138)', '(Liquid-3.085-4.131)', '(Liquid-4.372-5.5)', '(Drip-4.949-5.047)', '(Drip-5.279-5.458)', '(Generic impact sounds-8.819-9.142)', '(Drip-9.511-9.649)'\\ \hline

\textbf{Instruction:}Assess the style and elements of the rapping and music in this audio clip. Based on this, what genre of music might the DJ be playing on stage? \textbf{Output:} Given the presence of rapping and electronic music, the DJ is likely playing Hip Hop or Electronic Dance Music (EDM).& YRjUZjMPP-nA & Electronic music plays as a whoosh sound follows, creating a lively atmosphere for the DJ performing on stage. & '(Rapping-0.0-0.376)', '(Music-0.0-10.0)', '(Rapping-0.685-1.663)', '(Rapping-2.295-2.837)', '(Sound effect-2.423-5.222)', '(Sound effect-7.427-10.0)'\\ \hline

\bottomrule                                      
\end{tabular}
}
\caption{\small Examples of \texttt{CompA-R}}
\label{tab:aug_examples_prompt_train_2}
\end{table*}

\begin{table*}[]
\centering
\resizebox{\linewidth}{!}{
\begin{tabular}{p{0.33\linewidth}|p{0.17\linewidth}|p{0.15\linewidth}|p{0.35\linewidth}}
\hline \hline
\textbf{Instruction-Response Pairs} & \textbf{AudioSet ID} & \textbf{Caption} & \textbf{Timestamp Events} \\ \hline

\textbf{Instruction:}Identify the role of the crowd's continuous conversation during the woman's speech. How does it contribute to the atmosphere of the scene? \textbf{Output:} The crowd's constant chatter indicates an informal or relaxed environment, suggesting that while the woman's speech is central, other conversations are also ongoing in the background. & Y6fRYeClf5U4 & A woman delivers a speech while a crowd of people engage in conversation in an urban setting. & '(Crowd-0.0-10.0)', '(Wind-0.008-10.0)', '(Female speech, woman speaking-0.074-1.65)', '(Female speech, woman speaking-2.879-5.427)', '(Female speech, woman speaking-5.604-6.083)', '(Female speech, woman speaking-6.9-10.0)'\\ \hline

\textbf{Instruction:}Considering the presence of bird sounds and chainsaw noise, identify the probable location where this sound sequence is occurring. \textbf{Output:}The location is likely an outdoor area, possibly in a woodland or suburban setting where birds can be heard alongside human activity and power tools. & YbkG4M4TiXZg & A man speaks while a power tool revs up, indicating a speech event followed by the sound of a chainsaw. & '(Male speech, man speaking-0.0-0.268)', '(Chainsaw-0.0-10.0)', '(Male speech, man speaking-1.772-4.425)', '(Male speech, man speaking-5.008-8.118)', '(Bird vocalization, bird call, bird song-5.362-7.512)', '(Bird vocalization, bird call, bird song-8.244-8.709)', '(Bird vocalization, bird call, bird song-8.937-9.283)', '(Male speech, man speaking-9.661-10.0)'\\ \hline

\bottomrule                                      
\end{tabular}
}
\caption{\small Examples of \texttt{CompA-R}-\textbf{test}}
\label{tab:aug_examples_prompt_test_1}
\end{table*}

\begin{table*}[]
\centering
\resizebox{\linewidth}{!}{
\begin{tabular}{p{0.33\linewidth}|p{0.17\linewidth}|p{0.15\linewidth}|p{0.35\linewidth}}
\hline \hline
\textbf{Instruction-Response Pairs} & \textbf{AudioSet ID} & \textbf{Caption} & \textbf{Timestamp Events} \\ \hline

\textbf{Instruction:}Analyze the frequency and consistency of the ticking sound. What is the likely source of this sound, and what implication might it have on the depicted setting? \textbf{Output:}The steady ticking likely comes from a clock, which coupled with the soft music, suggests a relaxed, cozy domestic setting, perhaps aimed at unwinding or relaxation. & YCoBAR5Mbjys & The clock ticks steadily as soft music plays in the background, creating a calming atmosphere in a cozy living room. & '(Mechanisms-0.0-10.0)', '(Alarm clock-0.008-10.0)', '(Tick-0.386-0.583)', '(Tick-1.071-1.22)', '(Tick-1.764-1.906)', '(Tick-2.465-2.638)', '(Tick-3.197-3.331)', '(Tick-3.772-3.976)', '(Tick-4.346-4.48)', '(Tick-4.646-4.787)', '(Tick-5.087-5.22)', '(Tick-5.669-5.795)', '(Tick-6.031-6.15)', '(Tick-6.37-6.528)', '(Tick-6.724-6.795)', '(Tick-6.969-7.118)', '(Tick-7.386-7.614)', '(Tick-8.134-8.354)', '(Tick-8.882-9.094)', '(Tick-9.315-9.425)', '(Tick-9.575-9.685)'\\ \hline

\textbf{Instruction:}Identify the type of vocal music that is being depicted in the audio based on the presence of singing and beatboxing. \textbf{Output:}This audio resembles A Capella, where voices impersonate the sounds of instruments, including rhythms often mimicked through beatboxing. & Y6SvDRiIG2NY & A group of people sing and harmonize, creating vocal music with occasional beatboxing, in a room with a piano. & '(Male singing-0.0-6.594)', '(Music-0.0-10.0)', '(Mechanisms-0.0-10.0)', '(Breathing-7.064-8.314)', '(Breathing-8.911-10.0)', '(Male singing-9.713-10.0)'\\ \hline

\textbf{Instruction:}Based on the audio, ascertain the possible relationship between the gunfire sounds, artillery fire, and music. How does the sequencing and manner of these sounds contribute to the atmosphere of the scene? \textbf{Output:} The gunfire and artillery sounds likely serve as a ceremonial display, with the music adding to the grandeur and solemnity of a military parade.& YbJvOp4gmHBg & Gunshots and artillery fire echo through the air as music plays during a military parade at a raceway. & '(Music-0.0-10.0)', '(Generic impact sounds-0.166-0.307)', '(Artillery fire-0.32-0.704)', '(Generic impact sounds-0.781-0.948)', '(Generic impact sounds-1.063-1.165)', '(Generic impact sounds-1.524-1.677)', '(Generic impact sounds-2.625-2.881)', '(Artillery fire-3.035-3.521)', '(Generic impact sounds-3.611-3.777)', '(Generic impact sounds-4.213-4.43)', '(Generic impact sounds-5.096-5.262)', '(Artillery fire-5.288-5.762)', '(Generic impact sounds-5.89-6.095)', '(Generic impact sounds-6.479-6.812)', '(Generic impact sounds-6.94-7.106)', '(Artillery fire-7.222-7.606)', '(Generic impact sounds-8.207-8.425)', '(Artillery fire-8.476-8.988)', '(Generic impact sounds-9.206-9.385)', '(Generic impact sounds-9.654-9.795)'\\ \hline

\bottomrule                                      
\end{tabular}
}
\caption{\small Examples of \texttt{CompA-R}-\textbf{test}}
\label{tab:aug_examples_prompt_test_2}
\end{table*}

\end{document}